\documentclass[aps,pra,twocolumn,a4paper,showpacs,superscriptaddress,floatfix,10pt]{revtex4}
\usepackage[pdftex]{graphicx}
\usepackage{dcolumn}
\usepackage{bm}
\usepackage[usenames, dvipsnames]{color}
\usepackage{comment}

\usepackage[T1]{fontenc}

\usepackage{amsfonts}
\usepackage{amssymb}
\usepackage{amsmath}
\usepackage{amsthm}
\usepackage{mathrsfs}
\usepackage{multirow}

\usepackage{subfigure}
\usepackage{rotate}
\usepackage{dsfont}

\usepackage{array}
\newcolumntype{L}[1]{>{\raggedright\let\newline\\\arraybackslash\hspace{0pt}}m{#1}}
\newcolumntype{C}[1]{>{\centering\let\newline\\\arraybackslash\hspace{0pt}}m{#1}}
\newcolumntype{R}[1]{>{\raggedleft\let\newline\\\arraybackslash\hspace{0pt}}m{#1}}
\usepackage{tabularx}

\def\KeyWord#1{$\backslash$\IfColor{$\!\!$\textRed{#1}\textBlack}{#1}$\!\!$}

\usepackage{wasysym}

\newcommand{\be}{\begin{equation} }
\newcommand{\ee}{\end{equation} }
\newcommand{\ba}{\begin{eqnarray} }
\newcommand{\ea}{\end{eqnarray} }

\def\em{\it}

\newcommand{\bit}{\begin{itemize}}
\newcommand{\eit}{\end{itemize}}
\newcommand{\ben}{\begin{enumerate}}
\newcommand{\een}{\end{enumerate}}

\newcommand{\x}{\sigma_\mathrm{x}}
\newcommand{\y}{\sigma_\mathrm{y}}
\newcommand{\z}{\sigma_\mathrm{z}}

\renewcommand{\i}{\mathrm{i}}
\renewcommand{\d}{\mathrm{d}}
\newcommand{\e}{\mathrm{e}}

\newcommand{\Bn}{{\vec{n}}}
\newcommand{\Bm}{{\vec{m}}}
\newcommand{\Bk}{{\vec{k}}}
\newcommand{\Bl}{{\vec{l}}}
\newcommand{\Bw}{{\vec{\Omega}}}
\newcommand{\Bt}{{\vec{\theta}}}
\newcommand{\dBt}{{\vec{\delta\theta}}}

\def\bra#1{\langle#1|}
\def\ket#1{|#1\rangle}
\def\braket#1#2{\langle#1|#2\rangle}
\def\qexp#1#2{\bra{#2}#1\ket{#2}}
\def\cexp#1{\left<#1\right>}
\def\com#1#2{\left[#1,#2\right]}
\def\tr#1{\mathrm{tr}\left[#1\right]}
\def\pdev#1#2{\frac{\partial #1}{\partial #2}}
\def\dev#1#2{\frac{\d #1}{\d #2}}

\begin{document}

\title{Topological classification of quasi-periodically driven quantum systems}

\author{P. J. D. Crowley}
\email{philip.jd.crowley@gmail.com}
\affiliation{Department of Physics, Boston University, Boston, MA 02215, USA}

\author{I. Martin}
\affiliation{Materials Science Division, Argonne National Laboratory, Argonne, Illinois 60439, USA}

\author{A. Chandran}
\affiliation{Department of Physics, Boston University, Boston, MA 02215, USA}

\date{\today}

\begin{abstract}
Few level quantum systems driven by $n_\mathrm{f}$ incommensurate fundamental frequencies exhibit temporal analogues of non-interacting phenomena in $n_\mathrm{f}$ spatial dimensions, a consequence of the generalisation of Floquet theory in frequency space.
We organise the fundamental solutions of the frequency lattice model for $n_\mathrm{f}=2$ into a quasi-energy band structure and show that every band is classified by an integer Chern number.
In the trivial class, all bands have zero Chern number and the quasi-periodic dynamics is qualitatively similar to Floquet dynamics.
The topological class with non-zero Chern bands has dramatic dynamical signatures, including the pumping of energy from one drive to the other, chaotic sensitivity to initial conditions, and aperiodic time dynamics of expectation values.
The topological class is however unstable to generic perturbations due to exact level crossings in the quasi-energy spectrum.
Nevertheless, using the case study of a spin in a quasi-periodically varying magnetic field, we show that topological class can be realised at low frequencies as a pre-thermal phase, and at finite frequencies using counter-diabatic tools.

\end{abstract}

\maketitle

\section{Introduction}

External time-dependent drives are indispensable to a quantum mechanic. At weak amplitude, they probe linear response~\cite{forster2018hydrodynamic}, while at strong amplitude, they enable Hamiltonian engineering~\cite{bukov2015universal,meinert2016floquet,holthaus2015floquet,seetharam2018steady,yao2007optical,oka2009photovoltaic,inoue2010photoinduced,kitagawa2010topological,gu2011floquet,lindner2011floquet,kitagawa2011transport,jiang2011majorana,dora2012optically,lindner2013topological,delplace2013merging,katan2013modulated,iadecola2013materials,rudner2013anomalous,cayssol2013floquet,lababidi2014counter,goldman2014periodically,grushin2014floquet,kundu2014effective,asboth2014chiral,carpentier2015topological,wang2017effect} 

The frequency content of the drive determines the nature of the steady state in a few level quantum system. When the drive has a single fundamental frequency, the Floquet theorem guarantees that observables vary quasi-periodically in time~\cite{shirley1965solution,sambe1973steady}, while a stochastic drive leads to stochastic behaviour. 

Recent advances in the construction and control of long-lived coherent qubits in a variety of condensed matter and quantum optical systems allow access to the interesting intermediate regime where the drive has a finite number $n_\textrm{f}$ of incommensurate frequencies~\cite{cirac200438,jelezko2004observation,devoret2004superconducting,taylor2005fault,trauzettel2007spin,gali2009theory,blatt2012quantum,wendin2017quantum}. 
Despite the lack of periodicity, the Floquet formalism can be generalized by treating the phase angle associated with each incommensurate frequency as an independent variable.
The fundamental solutions of the Schr{\"o}dinger equation, the so-called quasi-energy states, then follow from the solutions of a tight-binding model in $n_\textrm{f}$ independent synthetic dimensions in frequency space~\cite{ho1983semiclassical,verdeny2016quasi,martin2017topological}.

\citet{martin2017topological} recently exploited the synthetic dimensions to engineer energy pumping in the adiabatic regime. Specifically, Ref.~\cite{martin2017topological} studied a spin-$1/2$ in a magnetic field composed of two incommensurate frequencies $\Bw =(\Omega_1 ,\Omega_2)$:
\begin{gather}
H_{\mathrm{CI}}(t) = \vec{B}(t) \cdot \vec{\sigma} \label{eq:Hintro} 
\end{gather}
where
\begin{align}
\vec{B}(t) &=
\begin{pmatrix}
\sin(\Omega_1 t + \theta_{01}) \\
\sin(\Omega_2 t +\theta_{02}) \\
m-\cos(\Omega_1 t +\theta_{01}) - \cos(\Omega_2 t + \theta_{02})
\end{pmatrix}
\nonumber
\end{align}
Interpreting $\Omega_1 t$ and $\Omega_2 t$ as momenta, $H_{\mathrm{CI}}$ is the momentum-space Hamiltonian of a two-dimensional Chern insulator (CI) for $0<|m|<2$~\cite{qi2006topological,bernevig2006quantum}.
The Hall response of the Chern insulator at weak electric field translates to the quantized pumping of energy between the drives in the spin problem.
In contrast, when $|m|>2$, the insulator has no Hall response and the spin dynamics qualitatively resemble that of the one tone case. 

Could a different choice of driving Hamiltonian produce more exotic dynamics of the driven spin? 
We present an exhaustive classification of the quasi-energy states of a $d$-level quantum system (qudit) driven by $n_\mathrm{f}=2$ incommensurate frequencies.
The generalized Floquet formalism (Sec.~\ref{sec:dynamics}) produces $d$ fundamental solutions of the Schr\"{o}dinger equation:
\begin{equation}
\begin{aligned}
\ket{\psi_j(t)} &=  \e^{- \i \epsilon_j(\Bt_0) t} \ket{\phi^j(\Bw t+\Bt_0)}
\end{aligned}
\end{equation}
where $j=1 \cdots d$, $\epsilon_j(\Bt_0)$ is a quasi-energy and $\ket{\phi^j(\Bw t+\Bt_0)}$ is the associated quasi-energy state, which is periodic in both of its arguments.
The initial drive phases, $\Bt_0\in [0, 2\pi)^2 $, define the \emph{Floquet zone}.
The quasi-energies and states can be organized into a two-dimensional \emph{quasi-energy band structure} with $d$ bands over the Floquet zone.
The dynamical classes of the driven qudit are indexed by the $d$ integer Chern numbers $C_j$ associated to the bands.
We refer to the class with all $C_j = 0$ as trivial and any other class as topological.

Remarkably, the dispersion of band $j$ is fixed by its Chern number:  \begin{equation}
\nabla_{\Bt_0} \epsilon_j = \frac{C_j}{2\pi}(-\Omega_2,\Omega_1).
\label{eq:grad_eps_intro}
\end{equation}
We derive this result in Sec.~\ref{sec:QETopo}.  
Heuristically, in the frequency space tight-binding model $\vec{\Omega}$ plays the role of an electric field which Stark localizes the quasi-energy states in the $\vec{\Omega}$ direction.
In the perpendicular direction $\vec{P} = (-\Omega_2,\Omega_1)$, the states are localized if $ C_j=0 $ and delocalized otherwise. 
Varying $\Bt_0$ twists the phase in the $\vec{P}$ direction; only the delocalized eigenstates respond and move along the electric field direction.
Eq.~\eqref{eq:grad_eps_intro} quantifies the change in the quasi-energy due to the component of the translation in the direction of the electric field. 

\begin{figure}
\begin{center}
\includegraphics[width=0.85\columnwidth]{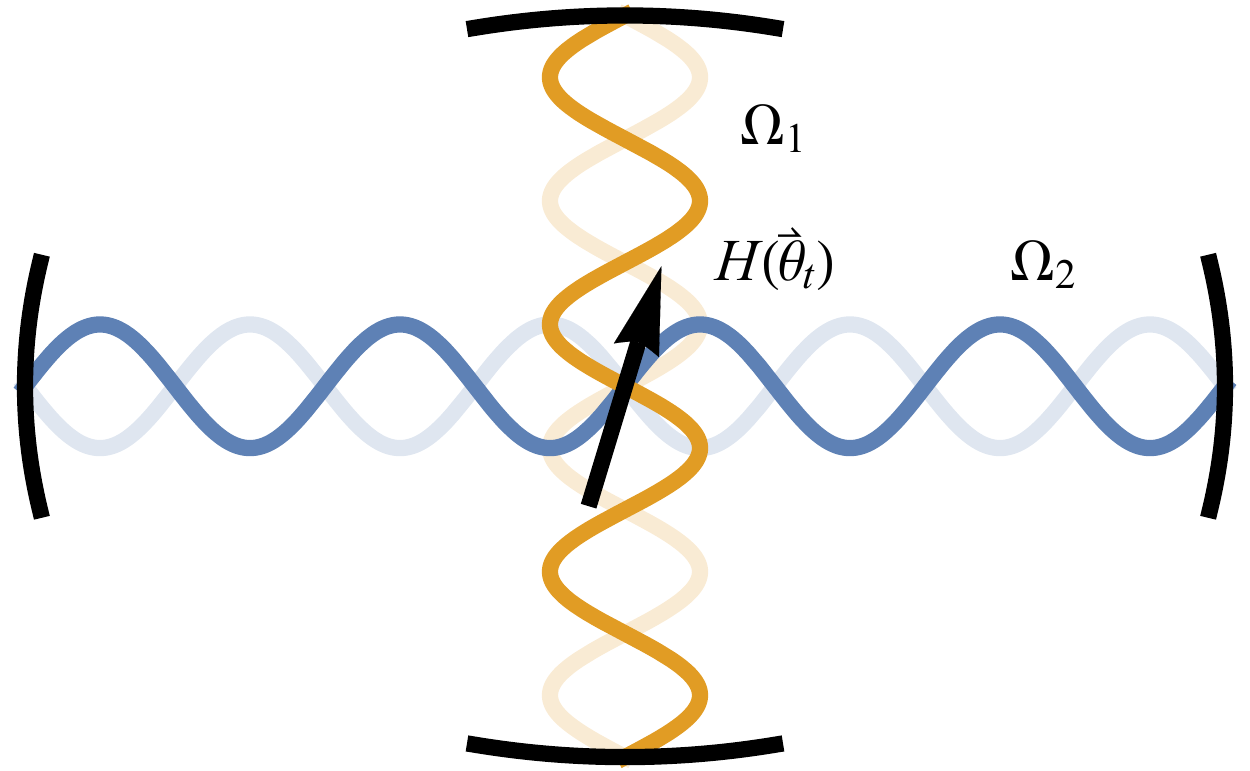}
\caption{
\emph{Two-tone driven quantum systems:} A $d$-level quantum system with Hamiltonian $H(\Bt_t)$ is driven by two classical cavity modes with frequencies $\Omega_1$ and $\Omega_2$.
}
\label{Fig:QPDriving}
\end{center}
\end{figure}

Eq.~\eqref{eq:grad_eps_intro} leads to stark differences in the dynamics starting from generic initial states for the topological and trivial dynamical classes (Sec.~\ref{sec:dynamicalproperties}).
The topological class is characterized by the pumping of energy between the drives, strong sensitivity to the initial phases of the drives and aperiodic dynamics of expectation values.
The latter two properties are properties of a \emph{quantum chaotic} qudit.
In contrast, the trivial class exhibits the same qualitative features as the periodically driven qudit: quasiperiodic dynamics with no energy pumping or chaos.
The sensitivity to the initial phase $\Bt_0$ is due to dephasing between the quasi-energy states which produces a linear in time divergence of expectation values. 
We believe that the distance diverges exponentially with a well-defined Lyapunov exponent if the external drive amplitudes are treated as dynamical variables. 

In the topological class, the quasi-energy band structures contain exact level crossings.
In the strict adiabatic limit the Chern indices are stable to Hamiltonian perturbations as they are inherited from a band insulator. 
At finite frequency, however, the level crossings are unstable to generic perturbations.
Using the Chern insulator model in~\eqref{eq:Hintro} as an example, we demonstrate that the topological class nevertheless controls (i) the pre-thermal dynamics in the vicinity of the adiabatic limit, and (ii) the dynamics for finite frequency drives with counter-diabatic terms (Sec.~\ref{sec:stability}).
We explicitly construct a counter-diabatic term $V(t)$ with finite spectral bandwidth that ensures that the quasi-energy states of
\begin{align} 
H_{\mathrm{CD}}(t)= H_{\mathrm{CI}}(t) + V(t)
\end{align} 
are given by the instantaneous eigenstates of $H_{\mathrm{CI}}(t)$.
Counter-diabatic (CD) terms stabilize the dynamical class of any adiabatic Hamiltonian $H(t)$ at finite drive frequency, and offer a route to realizing the topological class in the laboratory~\cite{del2013shortcuts,sels2017minimizing}.

Incommensurate external drives have been previously used to engineer the Anderson metal-insulator transition in kicked rotors~\cite{casati1989anderson,wang2009anderson}.
Previous studies have also discovered quasi-periodic and chaotic dynamical regimes in qudits driven by quasi-periodic sequences ~\cite{luck1988response,jauslin1991spectral,blekher1992floquet,jauslin1992generalized,barata2000formal,gentile2004pure}, and classified the quasi-energy states in terms of their monodromy~\cite{jauslin1991spectral,blekher1992floquet}.
Our classification in terms of band structures demonstrates completeness, generalizes to any number of tones, connects the dispersion to the Chern number and derives new dynamical properties of the dynamical classes. Using the counter-diabatic prescription, we also derive the first finite frequency quasi-periodically driven spin models in the topological class.

\section{Setup and Hamiltonian}
\label{sec:problemdef}

Consider a $d$-level quantum system driven by two ideal classical drives with fundamental frequencies $\Omega_1$ and $\Omega_2$ (Fig.~\ref{Fig:QPDriving}). 
Each drive is a $2\pi$-periodic function of its phase angle.
The phase angle $\theta_{ti}$ of drive $i$ increases linearly in time:
\begin{align}
  \theta_{ti} &= \Omega_i t + \theta_{0i}, \quad i= 1,2,
 \end{align} 
or more succinctly, $\Bt_t = \Bw t + \Bt_0$. The vector $\Bt_0$ sets the \emph{initial drive phases} at $t=0$. 

The Hamiltonian of the two-tone driven system is a $2\pi$-periodic function of each component of $\Bt_t$. It is therefore conveniently represented in Fourier series: 
\begin{equation}
H(\Bt_t ) = \sum_{\Bn} H_{\Bn} \e^{- \i \Bn\cdot\Bt_t },
\label{eq:H}
\end{equation}
with $\Bn = (n_1,n_2) \in \mathbb{Z}^2$.

The drive is quasi-periodic (or equivalently the frequencies $\Omega_1$ and $\Omega_2$ are incommensurate) if and only if $\Omega_1$ and $\Omega_2$ are rationally independent:
\begin{equation}
\Omega_2/\Omega_1 \equiv \beta \not\in \mathbb{Q}.
\end{equation}
In what follows, we use the terms quasi-periodic and incommensurate interchangeably and fix $\beta = (1+\sqrt{5})/2$ to be the golden ratio.

\subsubsection{Rational Approximation}
\label{sec:diophantine}

Determining whether two drive frequencies are quasi-periodically related requires infinite precision. 
We expect the finite time dynamics of the qudit to be insensitive to this property. 
The quasi-periodic case can therefore be approached through a limiting sequence of rationally (or commensurately) related drives:
\begin{equation}
p \Omega_1 = q \Omega_2.
\label{eq:rational_drive}
\end{equation}
Here $p,q$ are co-prime integers determined by the best rational approximations to $\beta$. The theory of Diophantine approximation defines the series of \emph{best rational approximations} $p/q$ to the irrational $\beta$ as the co-prime integers $p,q$ such that $|\beta-p/q|$ cannot be made smaller without increasing $q$. In the \emph{incommensurate limit}, in which $q$ is allowed to be arbitrarily large, one then finds $p/q \to \beta$~\cite{casselsintroduction,hindry2013diophantine}. The commensurate system is periodic with period $T = q T_1 = p T_2$ where $T_i = 2\pi /\Omega_i$. 
On time-scales $t\ll T$, we expect all observables to be the same as in the incommensurate limit, whereas on time-scales $t \gtrsim T$, the periodicity of the system becomes important and the dynamics are described by Floquet theory. 

Elementary results in the theory of Diophantine approximation state that~\cite{casselsintroduction,hindry2013diophantine}: (i) every irrational number has a unique infinite continued fraction expansion
\begin{equation}
\beta = a_0 + \frac{1}{a_1 + \frac{1}{a_2 + \frac{1}{a_3 + \ldots}}},
\label{eq:cfrac}
\end{equation}
and (ii) the best rational approximations $p_i/q_i$ to $\beta$ are given by truncating the continued fraction expansion at the $i$th level. For example, the best rational approximations to the golden ratio $\beta = (1+\sqrt{5})/2$ are given by $p_i/q_i = F_{i+2}/F_{i+1}$, where $F_i$ is the $i$th Fibonacci number. 

\section{Generalized Floquet theory and the quasi-energy band structure}
\label{sec:dynamics}

\begin{figure}
\begin{center}
\includegraphics[width=0.9\columnwidth]{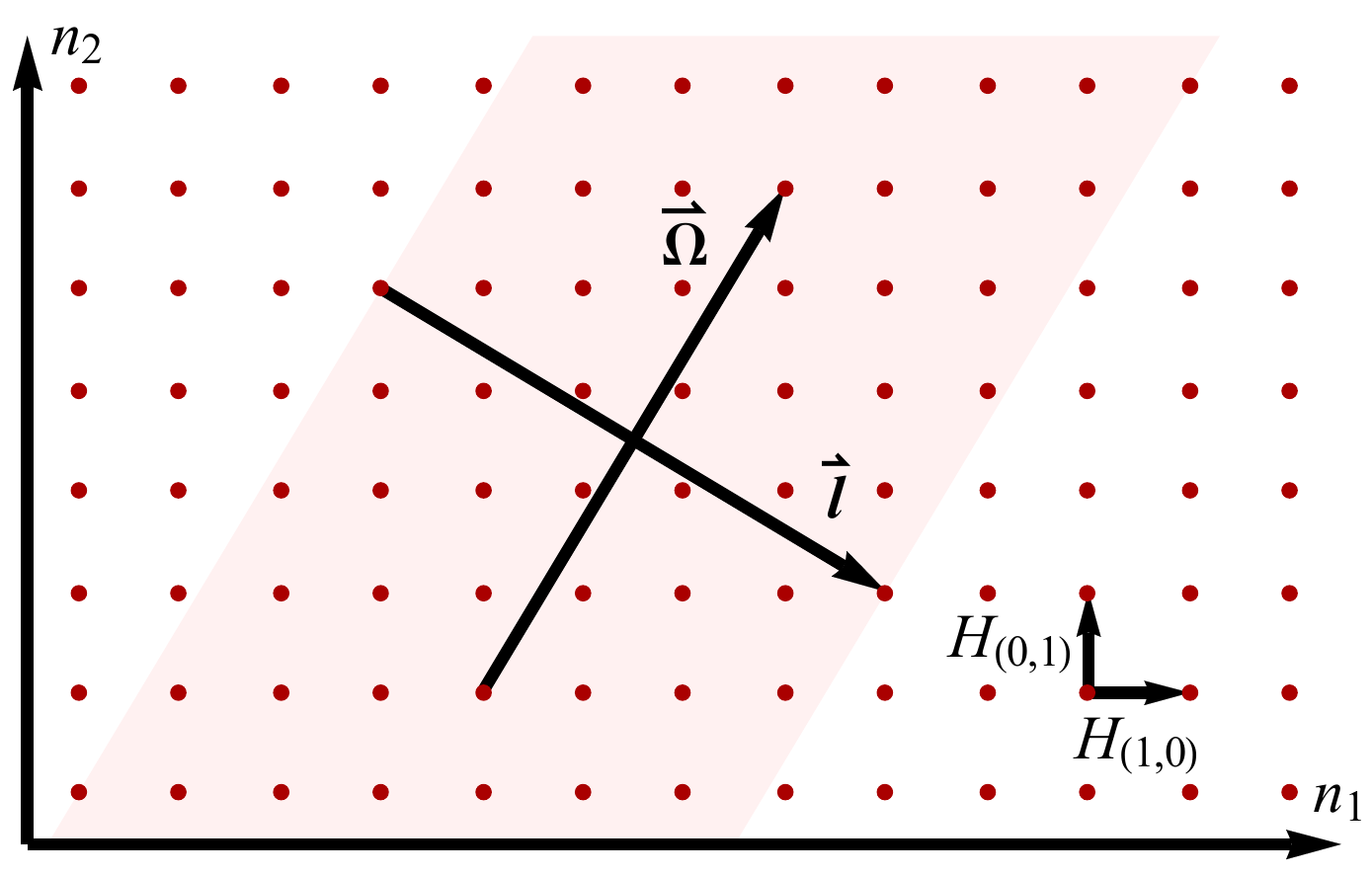}
\includegraphics[width=0.96\columnwidth]{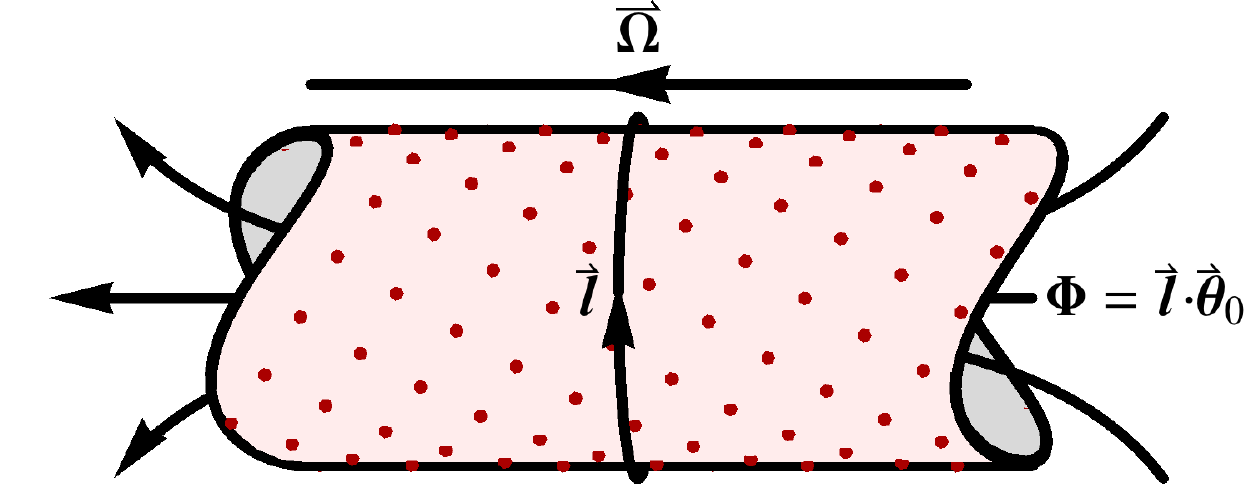}
\caption{
\emph{The frequency lattice:} In the incommensurate limit, solutions to Eq.~\eqref{eq:TDSE} are the solutions of a tight-binding model on an infinite two-dimensional lattice with an electric field $\Bw$ (upper panel). The Fourier components $H_\Bn$ couple sites $\vec{m}$ and $\vec{m} + \vec{n}$ for all $\vec{m}$. When the drive frequencies are commensurate, lattice sites separated by $\vec{l}$ are identified, and the shaded region compactifies into the cylinder shown in the lower panel. This cylinder encloses magnetic flux $\Phi$.
}
\label{Fig:FreqLattice}
\end{center}
\end{figure}
Let the state of the qudit at time $t$ be denoted by $\ket{\psi(t;\Bt_0)}$.
This state satisfies the Schr\"odinger equation:
\begin{equation}
\i \partial_t \ket{ \psi(t;\Bt_0)} = H(\Bw t + \Bt_0) \ket{\psi(t;\Bt_0)}.
\label{eq:TDSE}
\end{equation}
Below, we discuss the structure of the solutions to~\eqref{eq:TDSE} in the time and frequency domain by generalizing the Floquet formalism (see \cite{martin2017topological,verdeny2016quasi} for related treatments). We show that Fourier transforming~\eqref{eq:TDSE} yields a tight-binding model in frequency space in two synthetic dimensions (one for each rationally independent drive frequency). 
We use the spectrum of this tight-binding model to define the quasi-energy band structure.

\subsubsection{The quasi-energy operator and spectrum}
\label{sec:FreqLattice}

\begin{table}
  \renewcommand{\arraystretch}{1.45}
\begin{ruledtabular}
\begin{tabular}{ C{0.065\textwidth} | C{0.2\textwidth} | C{0.2\textwidth} }

& Time domain & Frequency domain  \\ 

\hline

$\Bn$ & Fourier index & Site index \\

$H_{\vec{0}}$ & Time averaged Hamiltonian & On-site potential \\

$H_\Bm$ & Fourier component of Hamiltonian& Hopping by vector $\Bm$ \\

$\ket{\tilde\phi^j_\Bn(\Bt_0)}$ & Fourier component of quasi-energy state & Quasi-energy state projected onto lattice site \\

$\Bw$ & Drive frequency vector & Electric field \\

$\Bt_0$ & Initial drive phase vector & Magnetic vector potential \\

$\beta$ & Ratio of drive frequencies $\beta=\Omega_2/\Omega_1$ & $\arctan \beta$ is the angle between $\hat{x}$ and $\Bw$ \\

$d$ & Hilbert space dimension of qudit & Number of orbitals per lattice site \\

\end{tabular}
\end{ruledtabular}
\caption{Dictionary relating quantities in the time and frequency domains.}
\label{tab:2domains}
\end{table}

\begin{figure*}
\begin{center}
\includegraphics[width=0.66\columnwidth]{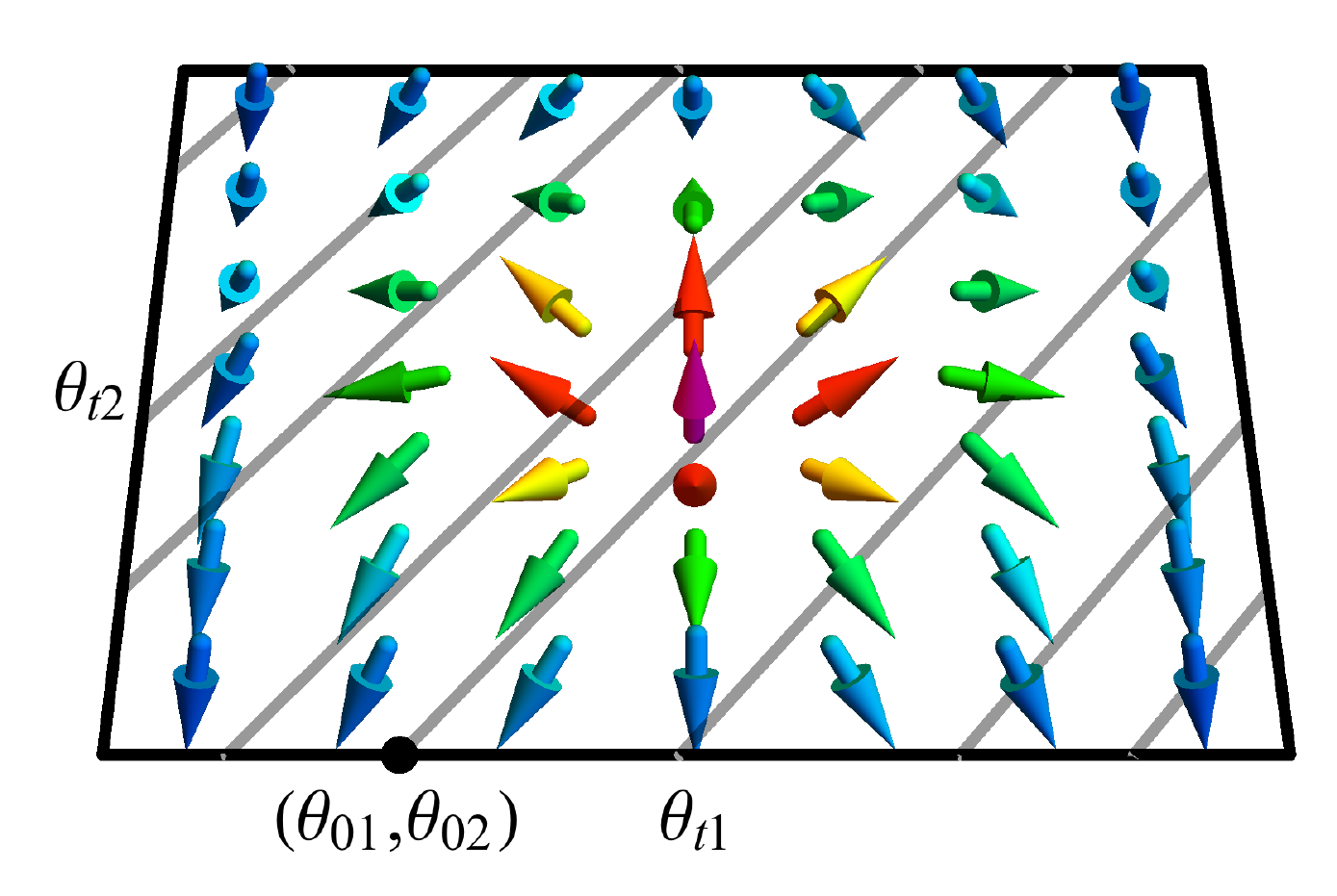}
\includegraphics[trim=-0.5cm -2cm 0 0,width=0.35\columnwidth]{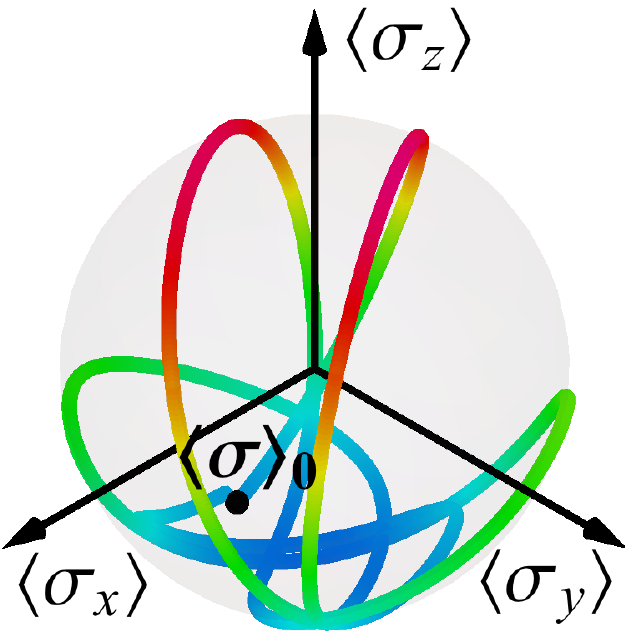}
\includegraphics[width=0.66\columnwidth]{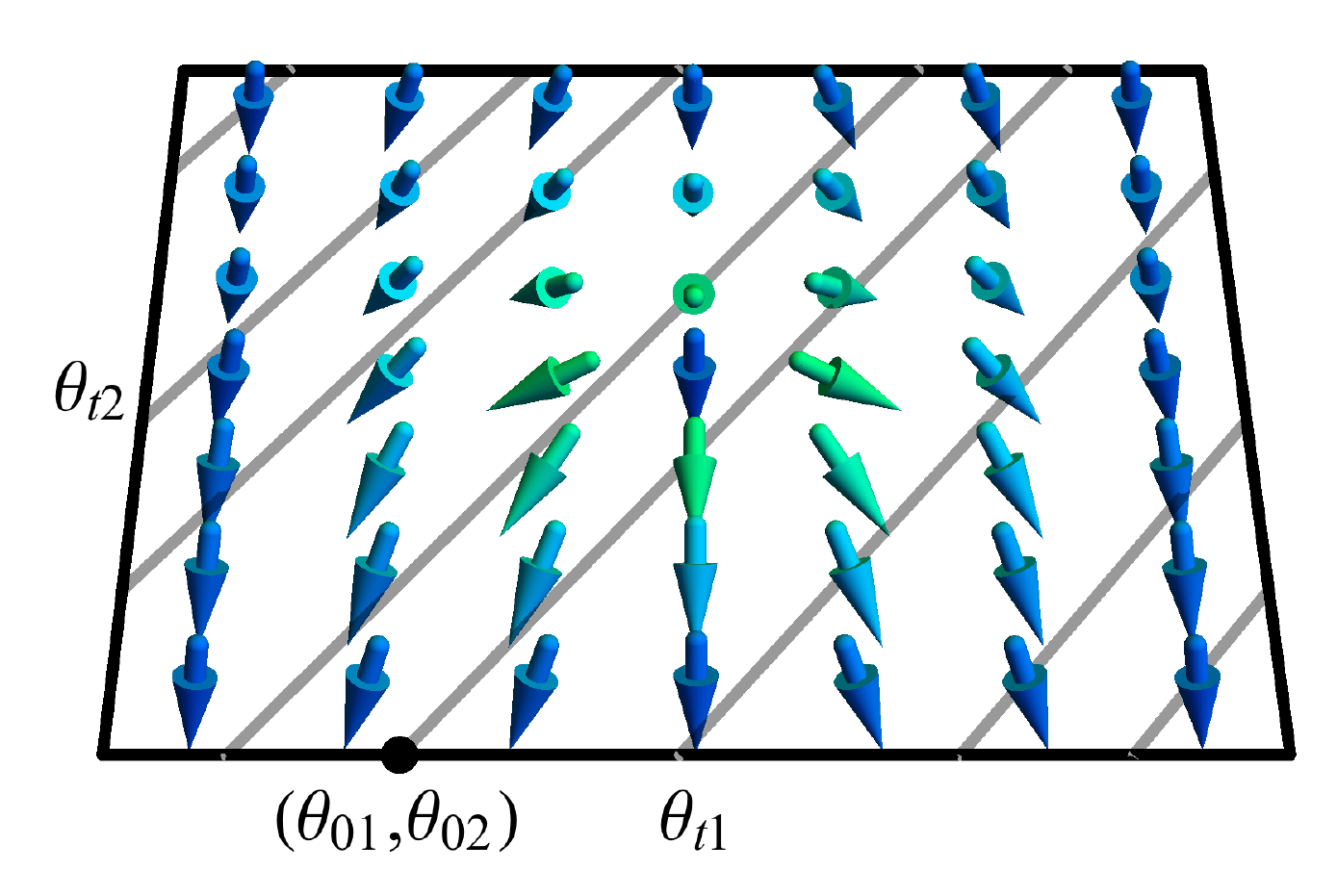}
\includegraphics[trim=-0.5cm -2cm 0 0,width=0.35\columnwidth]{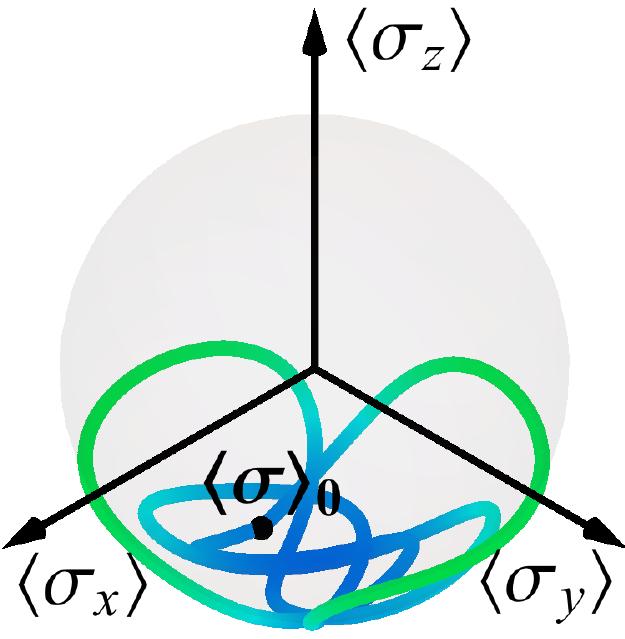}
\caption{
\emph{Visualizing the quasi-energy states:} The first and third panels show the Bloch vector $\qexp{\vec\sigma}{\phi^j(\Bt_t)}$ of a two-level system throughout the Floquet zone $\Bt_t \in [0,2\pi)^2$ for $j=1$. By~\eqref{eq:gauge_fix_phi}, the time evolution of a state starting from $\Bt_0$ corresponds to a straight path through the Floquet zone in the direction $\Bw$ (shown in grey for $0 < t < 5 T_2$). The second and fourth panels show the path of the Bloch vector on the Bloch sphere in the same interval. The color indicates the $z$-polarisation, from blue when $\cexp{\sigma_z}=-1$ to red when $\cexp{\sigma_z}=1$. The left (right) panels illustrate the topological (trivial) dynamical classes, and correspond to quasi-energy bands with $C_j=1(0)$. Data for $H_\mathrm{CD}$~\eqref{eq:topomodel} with $m=1(3)$ on the left(right), $\Bt_0=(-\pi/2,0)$.
}
\label{Fig:QEQEE}
\end{center}
\end{figure*}
Substituting the Fourier transform $\ket{\psi(t;\Bt_0)} = \int \d \omega \e^{-\i \omega t} \ket{\tilde\psi(\omega;\Bt_0)}$ into the Schr\"{o}dinger equation~\eqref{eq:TDSE}, we obtain
\begin{equation}
\omega\ket{\tilde\psi(\omega;\Bt_0)} = \sum_{\Bm  \in \mathbb{Z}^2} H_{\Bm} \e^{-i \Bm \cdot \Bt_0} \ket{\tilde\psi(\omega-\Bm\cdot\Bw;\Bt_0)}
\label{eq:FT_TDSE}
\end{equation}
where the Fourier coefficients $H_\Bm$ are defined in~\eqref{eq:H}. Eq.~\eqref{eq:FT_TDSE} only couples the frequencies,
\begin{equation}
\omega = \epsilon + \Bn\cdot\Bw,
\label{eq:FreqDec}
\end{equation}
for $\Bn \in \mathbb{Z}^2$ and fixed $\epsilon$. 
We can therefore find the \emph{fundamental solutions} $\ket{\tilde\phi(\omega;\Bt_0)}$ to~\eqref{eq:FT_TDSE} which are non-zero only for the frequencies~\eqref{eq:FreqDec}. 
Using the rational independence of $\Omega_1,\Omega_2$, we unambiguously label the Fourier components by $\vec{n}$ instead of $\omega$:
\begin{equation}
\ket{\tilde\phi_{\Bn}(\Bt_0)} \equiv \ket{\tilde\phi(\epsilon+ \Bn \cdot \Bw;\Bt_0)}.
\label{eq:labelscheme}
\end{equation}
There are multiple solutions of this form corresponding to different values of the quasi-energy $\epsilon(\Bt_0)$. Generic solutions to~\eqref{eq:FT_TDSE} are linear combinations of the fundamental solutions $\ket{\tilde\phi_{\Bn}(\Bt_0)}$ at different $\epsilon(\Bt_0)$. Combining~\eqref{eq:FT_TDSE},~\eqref{eq:FreqDec} and~\eqref{eq:labelscheme}, we obtain the eigenvalue equation:
\begin{align}
\epsilon(\Bt_0)\ket{\tilde\phi_\Bn(\Bt_0)} = \sum_{\Bm} &\left(  H_{\Bn-\Bm} \e^{-i (\Bn - \Bm )\cdot \Bt_0} \right. \nonumber \\
& \quad \quad \left. - \Bn\cdot\Bw \delta_{\Bn\Bm} \right)\ket{\tilde\phi_{\Bm}(\Bt_0)}.
\label{eq:Freqlattice}
\end{align}

We interpret $\Bn$ as the lattice sites of a two-dimensional hopping model in frequency space. Explicitly, we define: 
\begin{align}
\ket{\tilde\phi(\Bt_0)} &= \sum_{\Bn} \ket{\tilde\phi_\Bn(\Bt_0)} \otimes \ket{\Bn} \\
K(\Bt_0) &= \sum_{\Bn ,\Bm} \left[ H_{\Bn-\Bm} \e^{-i (\Bn - \Bm )\cdot \Bt_0} - \Bn\cdot\Bw \delta_{\Bn\Bm} \right] \otimes \ket{\Bn} \bra{\Bm}, \label{eq:K}
\end{align}
with $\braket{\Bn}{\Bm}=\delta_{\Bn\Bm}$. Then,~\eqref{eq:Freqlattice} becomes:
\begin{equation}
K(\Bt_0) \ket{\tilde\phi(\Bt_0)} = \epsilon (\Bt_0)\ket{\tilde\phi(\Bt_0)}.
\label{eq:FreqlatticeK}
\end{equation}
In analogy to Floquet theory, we refer to $\epsilon$, $K$ and $\ket{\tilde\phi(\Bt_0)}$ as the quasi-energy, the quasi-energy operator and the quasi-energy state respectively.
We also define the Floquet zone to be the torus generated by the initial drive phases $\Bt_0\in [0, 2\pi)^2$.

\subsubsection{A tight-binding model in frequency space}

We interpret the quasi-energy operator $K$ as the Hamiltonian of a two dimensional tight binding model using the dictionary in Table~\ref{tab:2domains}. $K$ consists of: (i) an on-site potential $H_{\vec{0}}$; (ii) hopping terms $H_\Bn$ which couple sites $\Bm$ to sites $\Bm+\Bn$; (iii) an electric field $\Bw$ in a non-lattice vector direction (in the electrostatic gauge); and (iv) a magnetic vector potential $\Bt_0$.
The bulk magnetic field is zero as $\Bt_0$ is spatially uniform. However, $\Bt_0$ encodes the twisted boundary conditions of the frequency lattice, as is most easily seen in the commensurate case.
For commensurate drives the sites $\Bn$ and $\Bn+\Bl$ correspond to the same frequency in~\eqref{eq:FreqDec}, where $\Bl = (-p,q)$ is a lattice vector perpendicular to $\Bw$. 
The sites $\Bn$ and $\Bn + \Bl$ should therefore be identified, which compactifies the two-dimensional lattice into a cylinder with circumference $|\Bl|$ (see Fig.~\ref{Fig:FreqLattice}). The cylinder encloses a magnetic flux 
\begin{equation}
\Phi = \oint \Bt_0 \cdot \d \vec{r} = \Bt_0 \cdot \Bl.
\label{eq:magflux}
\end{equation}

\subsubsection{The basis of quasi-energy states in the time domain}
\label{Sec:FloquetZone}
\label{Sec:TechnicalDetails}

\begin{figure}
\begin{center}
\includegraphics[width=\columnwidth]{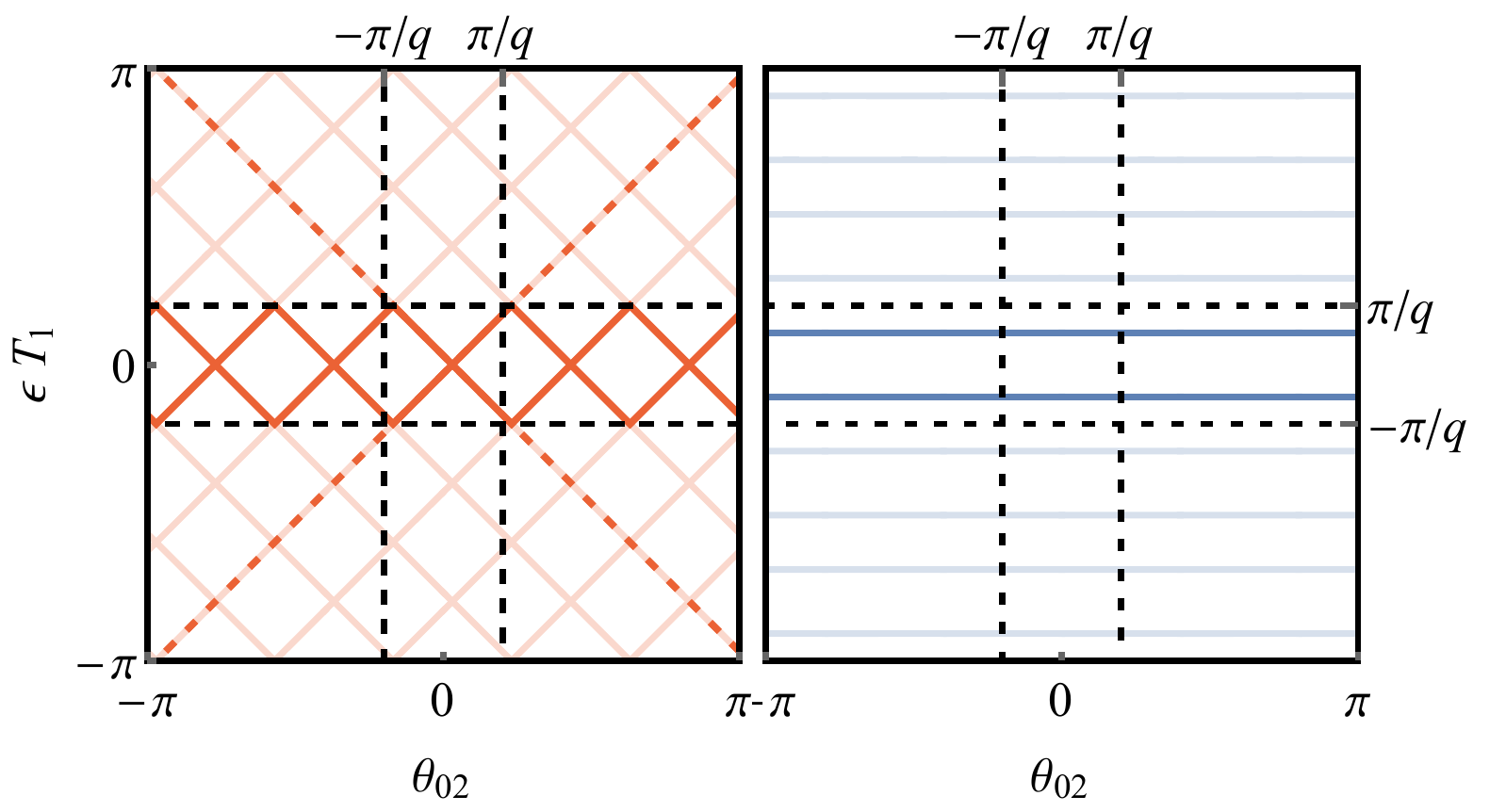}
\caption{
\emph{Band-structure of the quasi-energy operator:} A one-dimensional cut ($\theta_{01}=0$) of the two-dimensional band-structure of the quasi-energy operator $K$ for the topological (left) and trivial (right) classes of dynamics in the commensurate approximation. The unbounded spectrum is truncated to $\epsilon_j T_1 \in [-\pi,\pi]$. 
The solid dark bands are the reduced zone scheme: there is one band with constant positive gradient, and one with constant negative gradient. The dashed dark bands are in the extended zone scheme, while the light bands only appear in the repeated zone scheme. Data from $H_\mathrm{CD}$~\eqref{eq:topomodel} with $(p,q)=(8,5)$, and $m=1$ (left), and $m=3$ (right).}
\label{Fig:BandStructPlot}
\end{center}
\end{figure}

Each distinct solution to the Schr\"{o}dinger equation~\eqref{eq:TDSE} in the time domain identifies an equivalence class of quasi-energy states on the frequency lattice that are related by lattice translations.
This observation resolves the discrepancy between the infinite number of orthonormal solutions on the frequency lattice and the $d$ orthonormal solutions in the time domain.

The quasi-energy states in the time domain are obtained by inverse Fourier transform
\begin{equation}
\ket{\phi^j(t;\Bt_0)} = \sum_\Bn \mathrm{e}^{-\i \Bn \cdot \Bw t}\ket{\tilde\phi_\Bn^j(\Bt_0)},
\label{eq:QEdyn}
\end{equation}
where $j=1,\ldots d$ labels an orthonormal basis of solutions and $\ket{\tilde\phi_\Bn^j(\Bt_0)}$ is a representative element of the $j$th equivalence class in the frequency lattice. 
The corresponding solutions to the Schr\"{o}dinger equation~\eqref{eq:TDSE} are given by 
\begin{equation}
\ket{\psi^j(t)}=\mathrm{e}^{- \i \epsilon_j(\Bt_0) t} \ket{\phi^j(t;\Bt_0)}.
\label{eq:QEdyn2}
\end{equation}

To see that quasi-energy states related by lattice translations on the frequency lattice correspond to the same solution in the time domain, let $S_\Bm$ denote the translation by a frequency lattice vector $\Bm$:  $S_\Bm \ket{\Bn} = \ket{\Bn+\Bm}$. On conjugation by $S_\Bm$, $K$ is shifted by a constant 
\begin{equation}
S_\Bm K  S_\Bm^\dag = K + \Bm \cdot \Bw.
\label{eq:KShift}
\end{equation}
Hence for each quasi-energy state $\ket{\tilde\phi(\Bt_0)}$ with quasi-energy $\epsilon$, $\ket{\tilde\phi'(\Bt_0)} = S_\Bm \ket{\tilde\phi(\Bt_0)}$ is a quasi-energy state with $\epsilon' = \epsilon + \Bm \cdot \Bw$. It follows that:
\begin{equation}
\begin{aligned}
\mathrm{e}^{- \i \epsilon t} \ket{\phi(t;\Bt_0)}
 & = \e^{- \i \epsilon t} \sum_{\Bn}\e^{- \i \Bn \cdot \Bw t} \ket{\tilde\phi_\Bn(\Bt_0)} \\
& = \e^{- \i (\epsilon + \Bm \cdot \Bw )t} \sum_{\Bn}\e^{- \i \Bn \cdot \Bw t} \ket{\tilde\phi_{\Bn+\Bm}(\Bt_0)} \\
& = \e^{- \i \epsilon' t} \sum_{\Bn}\e^{- \i \Bn \cdot \Bw t} \ket{\tilde\phi_{\Bn}'(\Bt_0)} \\
& = \mathrm{e}^{- \i \epsilon' t} \ket{\phi'(t;\Bt_0)}.
\end{aligned}
\label{eq:equiv_bands}
\end{equation}

Generic solutions $\ket{\psi(t)}$ to the Schr\"{o}dinger equation~\eqref{eq:TDSE} are linear combinations of the quasi-energy states with their corresponding phases
\begin{equation}
\ket{\psi(t)} = \sum_j \alpha_j \e^{-\i \epsilon_j(\Bt_0) t} \ket{\phi^j(t;\Bt_0)}
\label{eq:genstate}
\end{equation}
for constant coefficients $\alpha_j \in \mathbb{C}$.

\subsubsection{Redundancy of time translations and phase shifts}
\label{sec:guagefix}
The Hamiltonian $H(\Bt)$ is invariant under the transformation $t \to t + \tau$, $\Bt_0 \to \Bt_0 - \Bw \tau$.
Thus, $ \ket{\phi^j(t;\Bt_0)}$ and $ \ket{\phi^j(t+\tau;\Bt_0-\Bw \tau )}$ are solutions to the Schr\"{o}dinger equation at the same quasi-energy $\epsilon_j$.
Choosing $\tau = -t$ we see that
\begin{equation}
\ket{\phi^j(t;\Bt_0)} \sim \ket{\phi^j(0; \Bt_0 + \Bw t)}.
\label{eq:gauge_eq}
\end{equation}
Above $\sim$ indicates equality up to multiplication by a phase. Eq.~\eqref{eq:gauge_eq} implies the information encoded by time evolution is also captured by a phase shift.

As the overall phase of a quasi-energy state is a gauge choice, i.e. physical observables evaluated in a quasi-energy state are invariant under the transformation $\ket{\phi^j(t;\Bt_0)} \rightarrow \e^{\i \Lambda_j(\Bt_0)} \ket{\phi^j(t;\Bt_0)}$, we are free to fix the phase in~\eqref{eq:gauge_eq} such that
\begin{equation}
\ket{\phi^j(t;\Bt_0)} = \ket{\phi^j(0; \Bt_0 + \Bw t)}.
\label{eq:gauge_fix_phi}
\end{equation}
Due to the equivalence of time evolution and phase shifts it is  then not necessary to keep track of $t$ and $\Bt_0$ separately. Henceforth we set $t=0$:
\begin{equation}
\ket{\phi^j(\Bt_0)} \equiv \ket{\phi^j(0;\Bt_0)} = \sum_\Bn \ket{\tilde\phi^j_\Bn(\Bt_0)}
\end{equation}
$\ket{\phi^j(\Bt_0)}$ is thus a periodic state defined over the toroidal Floquet zone. 

Though this gauge choice may not be smooth, the gauge invariant properties of $\ket{\phi^j(\Bt_0)}$ are smooth. For a two-level system, the gauge invariant properties of the state $\ket{\phi^j(\Bt_0)}$ are captured by the Bloch vector: $\qexp{\vec\sigma}{\phi^j(\Bt_0)}$, where $\vec\sigma$ is the vector of Pauli matrices. 
These Bloch vector fields are shown in Fig.~\ref{Fig:QEQEE} for the model $H_\mathrm{CI}$~\eqref{eq:Hintro}.

\subsubsection{Quasi-energy bands}
\label{sec:QEBands}

We promote the index $j$ from labelling the unique solutions at a specific values of $\Bt_0$ to a band index which labels a state for all $\Bt_0$. 

In the commensurate case with overall period $T = q T_1 = p T_2$ the symmetry~\eqref{eq:KShift} implies the band structure $\epsilon(\Bt_0)$ is invariant under the shift $\epsilon(\Bt_0) \to \epsilon'(\Bt_0)  = \epsilon(\Bt_0) + 2\pi/T$. By choosing the gauge~\eqref{eq:gauge_fix_phi} we work in the reduced zone scheme. The reduced zone scheme (pale solid lines in Fig.~\ref{Fig:BandStructPlot}) corresponds to choosing the states with quasi-energies $\epsilon(\Bt_0) \in [-\pi/T,\pi/T]$. These states lie within first `Brillouin zone' (between the horizontal dashed black lines in Fig.~\ref{Fig:BandStructPlot}). In this scheme the quasi-energy band structure is invariant under shifts in the time direction $\epsilon(\Bw t+\Bt_0)=\epsilon(\Bt_0)$, in the $\theta_{01}=0$ cut shown in Fig.~\ref{Fig:BandStructPlot} this invariance leads to the corresponding invariance of the quasi-energies under the shift $\theta_{02}\to \theta_{02}'=\theta_{02}+2\pi/q$. Note however, this symmetry of the quasi-energies is not a symmetry of the quasi-energy states $\ket{\phi^j(\Bt_0)}$. In the incommensurate limit $q \to \infty$ the reduced zone scheme is not well defined as the Brillouin zone $\epsilon \in [-\pi/T,\pi/T]$ collapses, however we will see that properties such as the quasi-energy gradient $\nabla_{\Bt_0} \epsilon(\Bt_0)$ remain well defined.

The reduced zone scheme is related to the alternative `extended zone scheme' by unfolding (dashed dark lines in Fig.~\ref{Fig:BandStructPlot}), this scheme leads to quasi-energies $\epsilon_j(\Bt_0)$ that are well defined in the quasi-periodic limit, but lacks the useful property $\epsilon(\Bw t+\Bt_0)=\epsilon(\Bt_0)$. In addition, the repeated zone scheme corresponds to considering the full set of bands (pale solid lines in Fig.~\ref{Fig:BandStructPlot}), 

Note that quasi-energies $\epsilon_j(\Bt_0)$ are not ordered by band index in general due to the possibility of exact band crossings (Fig.~\ref{Fig:BandStructPlot}).

\section{Topological classification of quasi-energy states}
\label{sec:QETopo}

The identification of a quasi-energy band structure allows us to use the familiar tools of momentum-space band theory to classify the bands. 
Treating the Floquet zone as the momentum-space Brillouin zone, we immediately see that each band should be characterized by an integer Chern number \cite{thouless1982quantized,bernevig2013topological}.
The Chern number $C_j$ of band $j$ is defined by equating $2\pi C_j$ to the Berry curvature of the quasi-energy states $\ket{\phi^j(\Bt_0)}$ integrated over the Floquet zone.

In static electronic systems, the dispersion is generally not constrained by the Chern number and researchers frequently work with modified Hamiltonians with completely flat energy dispersions~\cite{neupert2011fractional,mcgreevy2012wave,bernevig2013topological}.
Here, we derive the remarkable result that the gradient of the quasi-energy dispersion is fixed by the Chern number: 
\begin{equation}
\nabla_{\Bt_0} \epsilon_j(\Bt_0) = \frac{C_j}{2\pi}(-\Omega_2,\Omega_1).
\label{eq:grad_eps}
\end{equation}

\begin{figure}
\begin{center}
\includegraphics[width=\columnwidth]{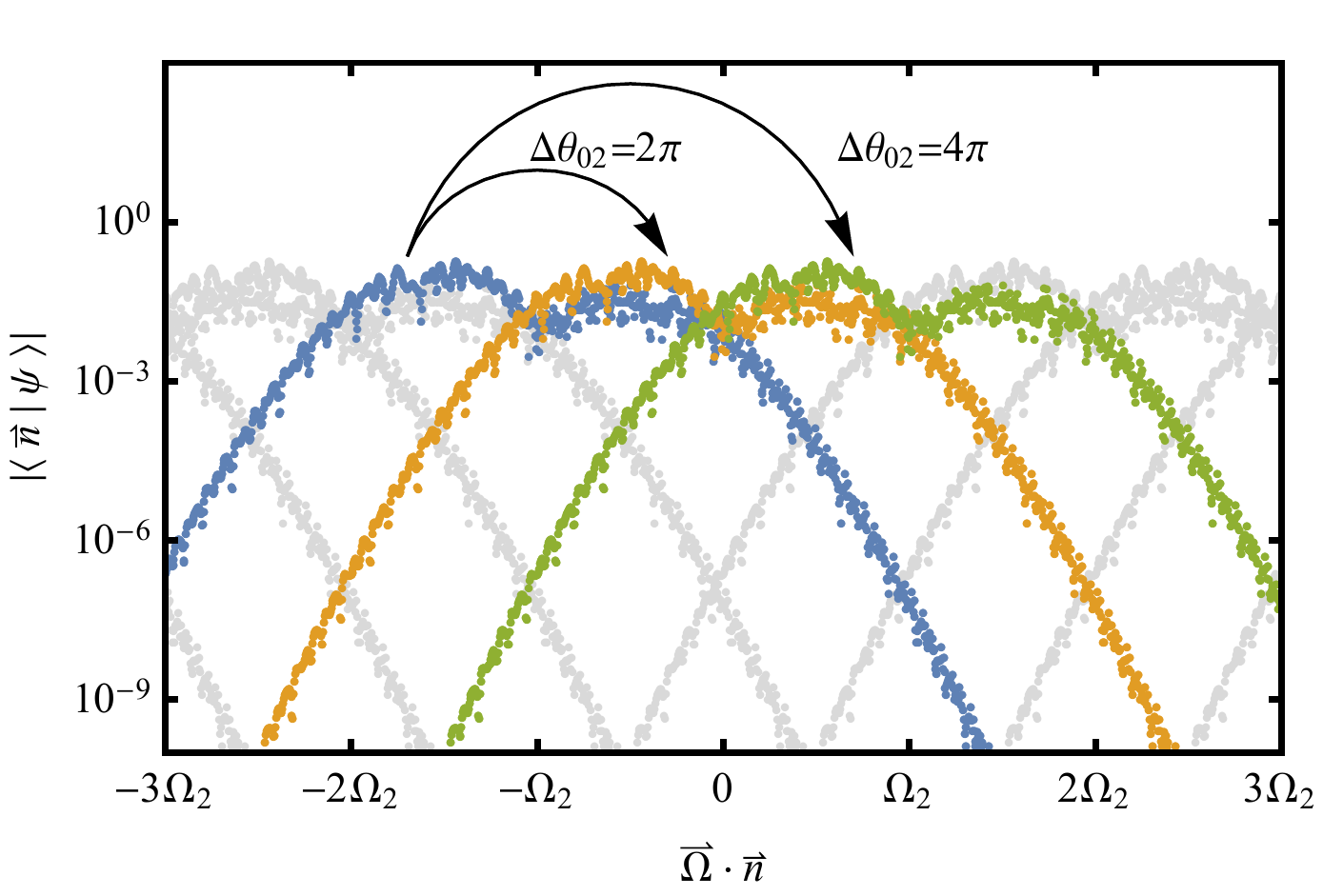}
\caption{
\emph{Shift of quasi-energy states on the frequency lattice with flux threading:} 
Amplitudes $|\left< \Bn | \psi \right> |$ of quasi-energy states belonging to a band with Chern number $C=1$ versus the electric potential energy $\Bw\cdot\Bn$ of the frequency lattice site $\vec{n}$. Threading a $2\pi$ flux through the frequency lattice cylinder increases the potential energy of the states and translates them along the electric field. The highlighted states separated by $\Delta \Bn = (1,0),(2,0)$ are related by flux changes of $\Delta \theta_{02} = 2\pi,4\pi$ respectively. Data from $H_\mathrm{CD}$~\eqref{eq:topomodel} with $(p,q)=(34,21)$, $\Omega_1 = 2\pi/20$ and $m=1$.
}
\label{Fig:Flux_Shifting}
\end{center}
\end{figure}

The origin of~\eqref{eq:grad_eps} lies in the response of quasi-energy states on the frequency lattice to flux threading. Consider varying $\Bt_0$ along the line $\theta_{01} =0$; Fig.~\ref{Fig:BandStructPlot} shows band structures along this path in the commensurate case.
An increase of $\theta_{02}$ by $2\pi/q$ corresponds to a $2 \pi$-increase of the magnetic flux $\Phi = \Bt_0 \cdot \Bl$ threading the frequency lattice. 
As a flux of $2\pi$ is gauge equivalent to a flux of zero, the quasi-energy spectrum at $\theta_{02}$ and $\theta_{02}+2\pi/q$ are identical. 
However, if we follow quasi-energy states as we increase $\theta_{02}$, we find that states may exchange positions with one another.
Fig.~\ref{Fig:BandStructPlot} shows the two qualitatively distinct possibilities for $d=2, q=5$.
In the left panel, half of the states in the spectrum (pale solid lines) are shifted up on increasing $\theta_{02}$ by $2\pi/q$, while the other half are shifted down. Thus, $\Delta \epsilon_j = \pm 2\pi/q T_1 = \pm \Omega_1 /q$.
In contrast, the spectrum is invariant under arbitrary changes of $\theta_{02}$ in the right panel. 
As the bands in the left (right) panel have $C_j = \pm 1(0)$, we see that $ \Delta \epsilon_j =  C_j \Omega_1/ q = C_j \Omega_1 \Delta \theta_{02}/2\pi$.
In the incommensurate limit, $q \to \infty$ and we obtain the gradient form in Eq.~\eqref{eq:grad_eps}.

Mathematically, the total change in quasi-energy of a band on increasing $\theta_{02}$ by $2\pi/q$ is given by:
\begin{equation}
\Delta \epsilon_j = \int_0^{2\pi/q} \d \theta_{02} \, \pdev{\epsilon_j}{\theta_{02}}.
\label{eq:Wdef}
\end{equation}
We pick a band labelling scheme such that $\ket{\phi^j(\Bt_0)}$ and $\epsilon_j(\Bt_0)$ are continuous functions of $\Bt_0$. From the eigenvalue equation~\eqref{eq:FreqlatticeK}, we obtain:
\begin{equation}
\nabla_{\Bt_0} \epsilon_j(\Bt_0) = \bra{\tilde\phi^j(\Bt_0)} \left( \nabla_{\Bt_0} K(\Bt_0) \right) \ket{\tilde\phi^j(\Bt_0)},
\end{equation}
As derived in App.~\ref{app:dedtheta}, elementary Fourier analysis yields
\begin{align}
\pdev{\epsilon_j}{\theta_{02}} & = \bra{\tilde\phi^j(\Bt_0)} \left( \partial_{\theta_{02}} K(\Bt_0) 
\right)
\ket{\tilde\phi^j(\Bt_0)} \nonumber
\\
& = \lim_{t \to \infty} \frac{1}{t} \int_0^{t} \d s \bra{\phi^j(\Bt_s)} \left( \partial_{\theta_{02}} H(\Bt_s) \right) \ket{\phi^j(\Bt_s)}.
\label{eq:dedtheta}
\end{align}
The double integral obtained from~\eqref{eq:dedtheta} and~\eqref{eq:Wdef} provides a uniformly weighted integration over the Floquet zone $0\leq\theta_1,\theta_2<2\pi$. Thus
\begin{equation}
\Delta \epsilon_j  = \frac{1}{2 \pi q}\int_\mathrm{FZ} \d^2 \Bt \bra{\phi^j(\Bt)} \left( \partial_{\theta_{2}} H(\Bt) \right) \ket{\phi^j(\Bt)}.
\end{equation}
Integrating by parts gives
\begin{align}
\Delta \epsilon_j  & = -\frac{1}{2 \pi q} \int_\mathrm{FZ}  \d^2 \theta \left[\bra{\partial_{\theta_{2}} \phi^j(\Bt)} H(\Bt)\ket{\phi^j(\Bt)} + \textrm{h.c.} \right].
\label{eq:WHpsi}
\end{align}
Next we use the relation
\begin{equation}
\i \Bw \cdot\ket{\nabla_{\Bt_0} \phi^j(\Bt_0)} = \left[H(\Bt_0)-\epsilon_j(\Bt_0)\right] \ket{\phi^j(\Bt_0)},
\label{eq:thetaTDSE}
\end{equation}
obtained by substituting~\eqref{eq:QEdyn2} into the Schr\"{o}dinger equation~\eqref{eq:TDSE}.
Substituting~\eqref{eq:thetaTDSE} into~\eqref{eq:WHpsi} yields the gauge invariant result
\begin{equation}
\Delta \epsilon_j = \int_0^{2\pi/q} \d \theta_{02} \, \pdev{\epsilon_j}{\theta_{02}} = \frac{\Omega_1 C_j}{q}.
\label{eq:WC}
\end{equation}
where $C_j$ is the Chern number:
\begin{equation}
C_j = \frac{1}{2 \pi \i} \int_\mathrm{FZ} \d^2 \theta \left[ \braket{\partial_{\theta_2}\phi^j}{\partial_{\theta_1}\phi^j} - \braket{\partial_{\theta_1}\phi^j}{\partial_{\theta_2}\phi^j} \right].
\label{eq:Chern}
\end{equation}
For a two-level system $C_j$ counts the integer number of topological solitons in the Bloch vector field $\qexp{\vec{\sigma}}{\phi^j(\Bt_0)}$. Examples are shown for the topological and trivial cases in Fig.~\ref{Fig:QEQEE}.

In the incommensurate limit, we require that the first derivative of the quasi-energy exists.
This allows for the identification
\begin{equation}
\pdev{\epsilon_j}{\theta_{02}} = \lim_{\Delta \theta_{02}  \to 0} \frac{\Delta \epsilon_j}{\Delta \theta_{02}} =  \lim_{q \to \infty} \frac{q \Delta \epsilon_j}{2 \pi}  = \frac{\Omega_1 C_j}{2 \pi}.
\label{eq:BSChern}
\end{equation}
Repeating the above derivation for an increase of $\theta_{01}$ by $2 \pi /p$ yields the full relation~\eqref{eq:grad_eps}.

When the Chern number of a band is non-zero, the quasi-energy states are translated by a lattice vector $\vec{m}$ on threading a flux of $2\pi$ through the frequency lattice cylinder (see Fig.~\ref{Fig:Flux_Shifting}).
The vector $\vec{m}$ can be uniquely determined from the change in quasi-energy $\Delta \epsilon_j$.
For example, increasing $\theta_{02}$ by $2 \pi$ increases the quasi-energy of band $j$ by $\Omega_1 C_j$. Using~\eqref{eq:KShift}, we equate this change to $\Bm \cdot \Bw$ to obtain $\Bm=(C_j,0)$. 

Finally, the quasi-energy states $\ket{\phi^j(\Bt_0)}$ form a complete basis. It follows that the sum of Chern numbers of all the bands is zero at every $q$:
\begin{align}
\sum_{j=1}^d C_j = 0
\end{align}

\subsection{Relation of the Chern number to monodromy}
\label{sec:monodromy}

Previous works~\cite{jauslin1991spectral,blekher1992floquet} have classified the quasi-energy states of incommensurately driven systems by their monodromy. 
As a trivial monodromy is equivalent to a trivial Chern number $C_j=0$~\cite{stone2009mathematics}, the two classifications are equivalent.
For completeness, we briefly discuss the equivalence below.

The quasi-energy states belonging to band $j$ have trivial monodromy if and only if there exists a smooth choice of gauge $\ket{\phi^j_\mathrm{M}(0,\theta_{02})} \sim \ket{\phi^j(0,\theta_{02})}$  such that the following relation holds:
\begin{equation}
U(T_1,0;0,\theta_{02})\ket{\phi^j_\mathrm{M}(0,\theta_{02})} = \mathrm{e}^{-\i \lambda T_1} \ket{\phi^j_\mathrm{M}(0,\theta_{02}+2\pi\beta)}.
\label{eq:monodromy}
\end{equation}
Above, $\sim$ indicates equality up to a $\theta_{02}$ dependent phase, $\lambda$ is a constant independent of $\theta_{02}$, and $U$ is the time evolution operator:
\begin{equation}
U(t',t;\Bt_0) = \mathcal{T} \exp \left[- \i \int_{t}^{t'} \d s H(\Bw s+ \Bt_0)\right].
\label{eq:Ut}
\end{equation}

Assume that~\eqref{eq:monodromy} holds. We use time evolution to smoothly extend the definition of $\ket{\phi^j_\mathrm{M}}$ to the full Floquet zone: 
\begin{equation}
\ket{\phi^j_\mathrm{M}(\Omega_1 t , \Omega_2 t + \theta_{02})} = \e^{\i \lambda t}U(t,0;0,\theta_{02})\ket{\phi^j_\mathrm{M}(0,\theta_{02})}.
\label{eq:extendmono}
\end{equation}
Using the definition~\eqref{eq:extendmono} then~\eqref{eq:monodromy} implies that $\ket{\phi^j_\mathrm{M}(2\pi , \Omega_2 T_1 + \theta_{02})} = \ket{\phi^j_\mathrm{M}(0 , \Omega_2 T_1 + \theta_{02})}$. Thus, $\ket{\phi^j_\mathrm{M}}$ is a smooth function of the Floquet zone. By Stokes theorem, the integrated Berry curvature in~\eqref{eq:Chern} is zero. Thus, $C_j=0$.

If $C_j=0$, then, $\epsilon_j$ is independent of $\Bt_0$, and the quasi-energy state gauge $\ket{\phi^j(0,\theta_2)}$ in the gauge~\eqref{eq:gauge_fix_phi} satisfies~\eqref{eq:monodromy} with $\lambda=\epsilon_j$. We show in App.~\ref{app:Monodromy} that this is a smooth gauge, and how to transform to it from any initial smooth gauge $\ket{\phi^j_\mathrm{S}(\Bt_0)}$ which can be trivially constructed. Thus $C_j=0$ implies monodromy.

\section{Dynamical signatures of the topological class}
\label{sec:dynamicalproperties}

\begin{table*}
  \renewcommand{\arraystretch}{1.45}
\begin{ruledtabular}
\begin{tabular}{ R{0.38\textwidth} || C{0.25\textwidth} | C{0.25\textwidth} }

& Trivial (all $C_j = 0$) & Topological (atleast one $C_j \neq 0$)  \\ 
\hline
Gradient of dispersion  & $\displaystyle \nabla_{\Bt_0} \epsilon_j = (0,0)$ & $\displaystyle \nabla_{\Bt_0} \epsilon_j = \frac{C}{2\pi}(-\Omega_2,\Omega_1)$. 
\\ 
Sensitivity to perturbation of $\Bt_0$ & Trajectories almost re-phase quasi-periodically & Trajectories diverge linearly
\\
Quasi-energy states in frequency domain & Localised & Delocalised
\\
Quasi-energy states in time domain & Sparse Fourier spectra & Dense Fourier spectra
\\
Frequency lattice response to flux threading &  Quasi-energy states unchanged & Quasi-energy states shift parallel to the electric field 
\\
Pump power of band $j$ & $\displaystyle P_j=0$ & $\displaystyle P_j= \frac{C_j}{2\pi}\Omega_1\Omega_2$
\\
Floquet operator converges as $U_i \to U(T_i;\Bt_0)$ & Yes & No
\\
Floquet Hamiltonian Exists & Yes & No 
\\
Time evolution of operator expectation values & Quasi-periodic evolution & Aperiodic evolution
\end{tabular}
\end{ruledtabular}
\caption{Properties of the two classes of dynamics for a quasi-periodically driven quantum system}
\label{tab:2phases}
\end{table*}

A quasi-energy band with a non-zero Chern number has striking dynamical consequences.
Qudits in the topological class pump energy between the drives, are sensitive to the initial phases and have operator expectation values with dense Fourier spectra. 
Qudits in the trivial class exhibit none of these properties; see Table~\ref{tab:2phases}.
Below, we derive these dynamical consequences and illustrate them with plots for the model discussed in Sec.~\ref{sec:CD}.

\subsection{Energy pumping}
\label{sec:pumppump}

\begin{figure}
\begin{center}
\includegraphics[width=0.96\columnwidth]{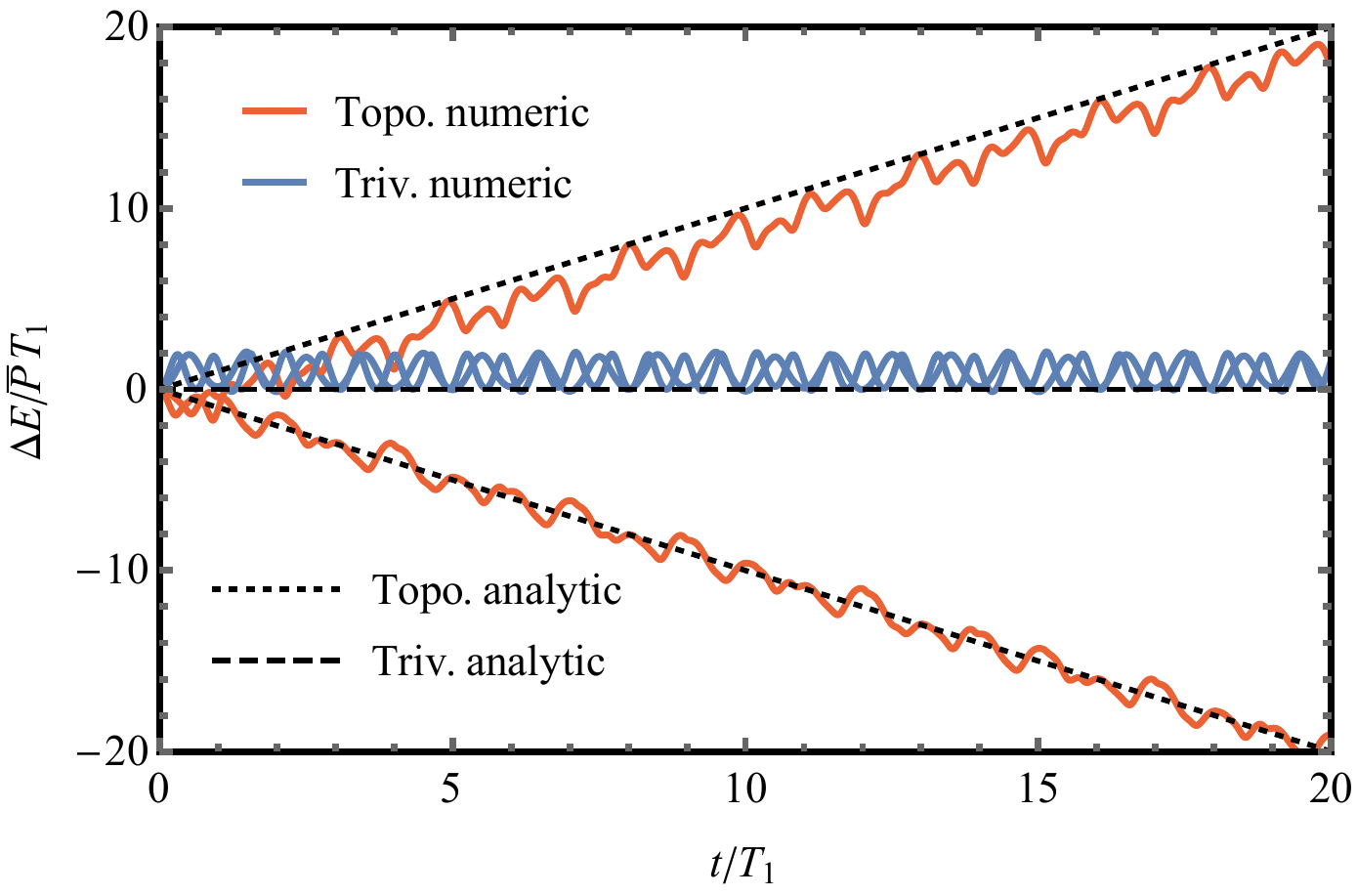}
\caption{
\emph{Energy pumping in quasi-energy states:} The scaled energy transfer between the two drives as a function of time in the topological (red) and trivial (blue) case for a two-level system prepared in a quasi-energy state. Asymptotically, the numerical curves are described by the relation: $\Delta E = P_j t$ (black lines). Data from $H_\mathrm{CD}$~\eqref{eq:topomodel} with $m = 1$ (red) and $m=3$ (blue).
}
\label{Fig:PumpingPlot}
\end{center}
\end{figure}
Ref.~\cite{martin2017topological} used an analogy with lattice Chern insulators to argue for quantized energy pumping in quasi-energy states in the adiabatic limit.
In the adiabatic limit, the electric field $\vec{\Omega}$ in the frequency lattice is weak (Table~\ref{tab:2domains}).
Suppose the model on the frequency lattice at $\vec{\Omega}=0$ is a Chern insulator.
At weak fields, the insulator exhibits the quantum Hall effect, that is, each eigenstate of the frequency lattice carries a quantised current perpendicular to $\vec{\Omega}$.  
As the components of the site label $\vec{n}=(n_1, n_2)$ on the frequency lattice equal the number of photons in the two drives (up to some arbitrary offset), the Hall effect leads to a quantized rate of transfer of energy between the two drives.

Below, we generalize the argument in Ref.~\cite{martin2017topological} to finite $\Bw$ and show that quantized energy pumping is a dynamical signature of quasi-energy states in the topological class of dynamics (see Fig.~\ref{Fig:PumpingPlot}).

The work done by the second drive up to a time $t$ on a system initially prepared in the quasi-energy-state is given by
\begin{equation}
\Delta E_j^{(2)}(t) = \int_0^{t} \d s \bra{\phi^j(\Bt_s)} \Omega_2 \partial_{\theta_{02}} H(\Bt_s)\ket{\phi^j(\Bt_s)}.
\label{eq:deltaE}
\end{equation}
The mean rate of work done by the second drive is then:
\begin{equation}
P_j^{(2)} = \lim_{t \to \infty} \frac{\Delta E_j^{(2)}(t)}{t}.
\label{eq:P2}
\end{equation}
As the qudit can only contain a finite amount of energy, the rate of work done by each of the two drives on the system must be equal and opposite at long times $P_j^{(1)}=-P_j^{(2)}$.
The system therefore behaves as an energy pump with power $P_j^{(1)}$.

Using Eq.~\eqref{eq:dedtheta}, we find that the pump power is set by the gradient of the quasi-energy dispersion:
\begin{equation}
P_j^{(2)}  = \Omega_2\frac{\partial \epsilon_j }{\partial\theta_{02}}.
\label{eq:Poweri}
\end{equation}
Eq.~\eqref{eq:grad_eps} then provides our result of quantized pumping in the topological class:
\begin{equation}
P_j^{(2)}  = - P_j^{(1)}  = \frac{C_j}{2 \pi}\Omega_1\Omega_2 .
\label{eq:pumppower}
\end{equation}

Generic initial states~\eqref{eq:genstate} also pump energy between the drives. Assuming the quasi-energy spectrum is non-degenerate, the contribution of cross terms averages to zero, and the pump power is:
\begin{equation}
P_\psi^{(n)} = \sum_j |\alpha_j|^2 P_j^{(n)}, \qquad n=1,2.
\end{equation}
We see that $0 \leq |P_\psi^{(n)}| \leq  \max_j|P_j^{(n)}|$.
Although the pump power is generically not quantized, it is non-zero except for a measure zero set of states. 

In the Chern insulator analogy, the transverse Hall current evaluated in quasi-energy states is the photon flux between the drives $\partial_t \cexp{\Bn}$. Using the results derived above:
\begin{equation}
\langle\vec{J}\rangle  = \partial_t \cexp{\Bn} = (P_j^{(1)}/\Omega_1,P_j^{(2)}/\Omega_2) = \frac{C_j}{2\pi}(-\Omega_2,\Omega_1),
\end{equation}
We thus recover the quantum Hall effect $\sigma_{xy} = |\Bw|/|\langle\vec{J}\rangle |=C_j/2\pi$ in natural units ($e=1,\hbar=1$).

\subsection{Divergence of trajectories}
\label{sec:divergence}

\begin{figure}
\begin{center}
\includegraphics[width=0.96\columnwidth]{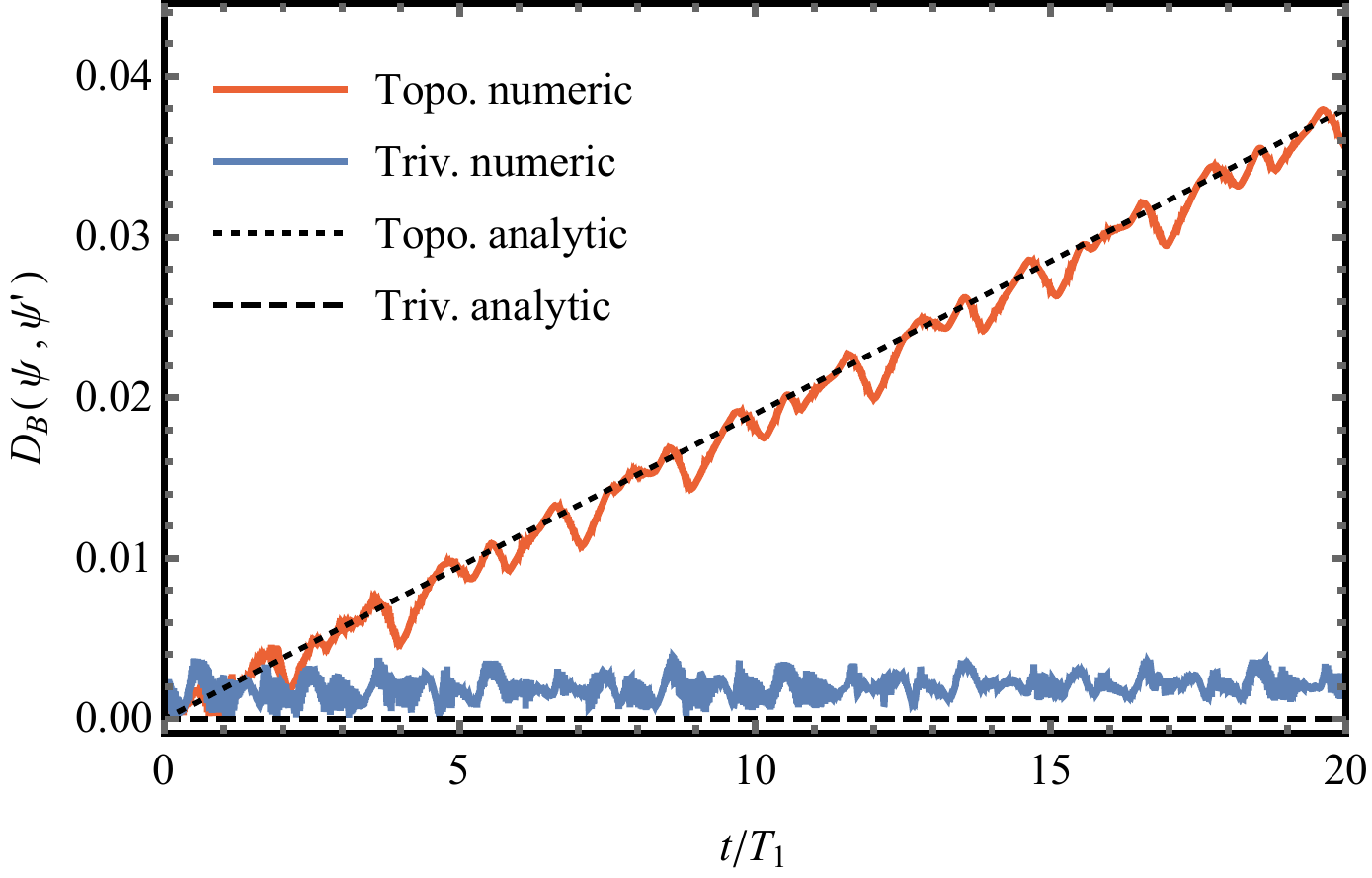}
\caption{
\emph{Divergence of trajectories:} In the topological dynamical class, trajectories diverge (red), while the trajectories in the trivial class do not (blue). The asymptotic behaviour in Eq.~\eqref{eq:errorgrowth} is shown in black.  Data from $H_\mathrm{CD}$~\eqref{eq:topomodel} with $m = 1$ (red) and $m=3$ (blue) for an initial state with $|\alpha_1| = |\alpha_2| = 1/\sqrt{2}$. 
}
\label{Fig:Divergence}
\end{center}
\end{figure}

Consider two time evolutions starting from the same initial state $\ket{\psi_0}$ but slightly different initial drive phases, $\Bt_0$ and $\Bt_0 + \delta \Bt$.
We show the trajectories of the perturbed and unperturbed system asymptotically diverge only in the topological case (see Fig~\ref{Fig:Divergence}).

The origin of the divergence between trajectories is dephasing in the quasi-energy basis. A quasi-energy state prepared with initial drive phase vector $\Bt_0$ evolves as
\begin{equation}
\begin{aligned}
U(t,0;\Bt_0)\ket{\phi^j(\Bt_0)} &= \mathrm{e}^{-\i \epsilon_j(\Bt_0) t}\ket{\phi^j(\Bt_t)} 
\end{aligned}
\end{equation}
where $U$ is the time evolution operator~\eqref{eq:Ut}.
If we choose a smooth gauge for the states $\ket{\phi^j(\Bt_0)}$ over the patch $\Bt \in \Bt_0 + s \Bw + r \delta \Bt$ for $0\leq s \leq t$, $0\leq r \leq 1$, we can expand the time evolution starting from $\Bt_0 + \delta \Bt$ to leading order in $\delta \Bt$. At leading order, the contribution from expanding $\ket{\phi^j(\Bt_t+\delta \Bt)}$ is $O(t^0\delta \theta)$, while the term from expanding the phasor $\mathrm{e}^{-\i \epsilon_j(\Bt_0+\delta \Bt) t}$ is $O(t^1\delta \theta)$. In the limit of small $\delta \Bt$ and large $t$, therefore we need only consider the second contribution.

After a time $t$, the phase difference $\mathrm{e}^{\i \eta}$ between the states with initial phase vector difference $\delta \Bt$ is 
\begin{equation}
\begin{aligned}
\eta &= \left( \epsilon_j(\Bt_0+\delta\Bt) - \epsilon_j(\Bt_0) \right) t + O(\delta \Bt^2) \\
& = t \delta \Bt \cdot \nabla_{\Bt_0} \epsilon_j \, + O(\delta \Bt^2) \\
& = \frac{C_j |\Bw|\sin\alpha}{2\pi} t |\delta \Bt| + O(\delta \Bt^2),
\end{aligned}
\label{eq:dephased}
\end{equation}
where $\alpha$ is the angle between $\Bw$ and $\dBt$, and we have used~\eqref{eq:grad_eps}~\footnote{We note that $A= \sin \alpha |\Bw||\dBt|t$ is the area in the Floquet zone enclosed by the two perturbed trajectories between the initial perturbation and measurement at time $t$. Generalising the argument here one finds that the phase difference between two perturbed quasi-energy state trajectories is given by $\eta = \frac{A C_j}{2 \pi}$}.

The global phase in~\eqref{eq:dephased} is unobservable in pure quasi-energy states. As $\eta$ depends on the band index,~\eqref{eq:dephased} leads to dephasing in the quasi-energy state basis for generic starting states.
To see how the dephasing leads to the divergence of trajectories, consider the evolution from the initial state $\ket{\psi_0}$ with and without the perturbation:
\begin{equation}
\begin{aligned}
\ket{\psi_t} &= U(t,0;\Bt_0)\ket{\psi_0},\\
\ket{\psi_t'} &= U(t,0;\Bt_0+\dBt)\ket{\psi_0}.
\end{aligned}
\label{eq:psi1psi2}
\end{equation}
For concreteness we characterise the distance between the two states using the Bures angle
\begin{equation}
D_\mathrm{B}(\psi,\psi') = \arccos \left|\braket{\psi}{\psi'}\right|,
\label{eq:BuresAngle}
\end{equation}
The Bures angle is a distance measure on quantum states~\footnote{Formally this measure is a statistical distance. This entails that the Bures angle has the properties: non-negativity $D_\mathrm{B}(\psi,\psi')>0$; symmetry $D_\mathrm{B}(\psi,\psi')=D_\mathrm{B}(\psi',\psi)$; identity of indiscernibles and $D_\mathrm{B}(\psi,\psi)=0$; and the triangle inequality $D_\mathrm{B}(\psi,\psi') \leq D_\mathrm{B}(\psi,\psi'') + D_\mathrm{B}(\psi'',\psi')$.} and bounds the discriminability of the two states using \emph{any} operator  $A$ via the bound
\begin{equation}
\left|\qexp{A}{\psi}-\qexp{A}{\psi'}\right|\leq 2|A|\sin \left[D_\mathrm{B}(\psi,\psi')\right]
\end{equation}
where the operator norm $|A|$ is the magnitude of the leading eigenvalue of $A$. 

For the trivial class of dynamics, the distance $D_\mathrm{B}(\psi_t,\psi_t')$ varies quasi-periodically in time and does not grow asymptotically. In contrast, for the topological class, $D_\mathrm{B}$ generically grows linearly in time, before saturating at long times to its maximal value $\max D_\mathrm{B} = \pi / 2$.
These results follow from the relation:
\begin{equation}
\lim_{t \to \infty} \lim_{|\delta \Bt| \to 0} \frac{D_\mathrm{B}(\psi,\psi')}{t|\delta \Bt|} = \frac{|\Bw|\,\sigma(C)\,\sin\alpha}{2\pi}.
\label{eq:errorgrowth}
\end{equation}
Appendix~\ref{app:divergence} contains the derivation. Above $\sigma(C)$ is the standard deviation of the Chern number in the initial state~\eqref{eq:genstate}:
\begin{equation}
\sigma^2(C) = \sum_j C_j^2 |\alpha_j|^2 - \left(\sum_j C_j |\alpha_j|^2 \right)^2.
\label{eq:sigmaC}
\end{equation}

\subsubsection{Convergence of Floquet unitaries}

We have shown that a perturbation to the initial conditions leads to a separation of trajectories for topological dynamics. A perturbation to the drive frequencies $\Omega_1,\Omega_2$ can be interpreted as many infinitesimal perturbations to the drive phases. Thus, using the same approach one can show an additional technical consequence of topology: in the trivial class of dynamics this leads to a convergence of the commensurate Floquet unitaries to the incommensurate time evolution operator
\begin{equation}
\lim_{i \to \infty} |U(q_i T_1, 0;\Bt_0) - U_i| = 0
\label{eq:UiUt}
\end{equation}
whereas in the topological case it does not. Here $U(q_i T_1, 0;\Bt_0)$ is the time evolution operator~\eqref{eq:Ut} in the incommensurate limit ($\Omega_2/\Omega_1 = \beta$), and whereas $U_i$ is the Floquet unitary of the commensurate approximation, found by integrating over the same period with frequencies $\Bw_i' = (\Omega_1,\Omega_1p_i/q_i)$
\begin{equation}
U_i = \mathcal{T} \exp \left[- \i \int_{0}^{q_i T_1} \d s H(\Bw_i' s+ \Bt_0)\right].
\label{eq:CommFloqOp}
\end{equation}
This can be understood as a perturbation to the second frequency $\Delta \Omega_2 = (p/q-\beta)\Omega_1 \sim 1/q^{2}$ to the second frequency. For trivial dynamics this convergence can be seen numerically via the corollary of~\eqref{eq:UiUt}
\begin{equation}
\lim_{i \to \infty} \left| U_{i} - \left(  U_{i-1} \right)^{a_i} U_{i-2}\right| = 0
\label{eq:UiUiUi_triv}
\end{equation}
where $a_i$ are the partial quotients defined via the continued fraction expansion of $\beta$~\eqref{eq:cfrac}. Eq.~\eqref{eq:UiUiUi_triv} follows composing the unitaries corresponding to smaller commensurate periods to approximate one of a larger commensurate period and uses the result of Diophantine approximation that $q_i = a_i q_{i-1}+ q_{i-2}$. The two terms in~\eqref{eq:UiUiUi_triv} correspond to two different closed paths through the Floquet zone, the limit converges if the small difference to the phase angles $\Bt_t$ between the two paths are inconsequential.

In contrast when accounting for topology we find that due to the effects discussed in Sec.~\ref{sec:divergence} the two trajectories accrue phase differently and there is a correction to the phase
\begin{equation}
\lim_{i \to \infty} \left| \left( U_{i} - (-1)^{a_i C_j} \left(  U_{i-1} \right)^{a_i} U_{i-2} \right)\ket{\phi^j(\Bt_0)} \right| = 0.
\label{eq:UiUiUi}
\end{equation}
The convergence relation has an additional sign $(-1)^{a_i C_j}$ which depends on the Chern number $C_j$ of each quasi-energy states subspace. This topological correction to the composition rule of the Floquet unitaries is derived in Appendix~\ref{app:FloqU}.

\subsection{Delocalisation on the frequency lattice and aperiodicity of observables}
\label{sec:deloc}

\begin{figure}
\begin{center}
\includegraphics[width=\columnwidth]{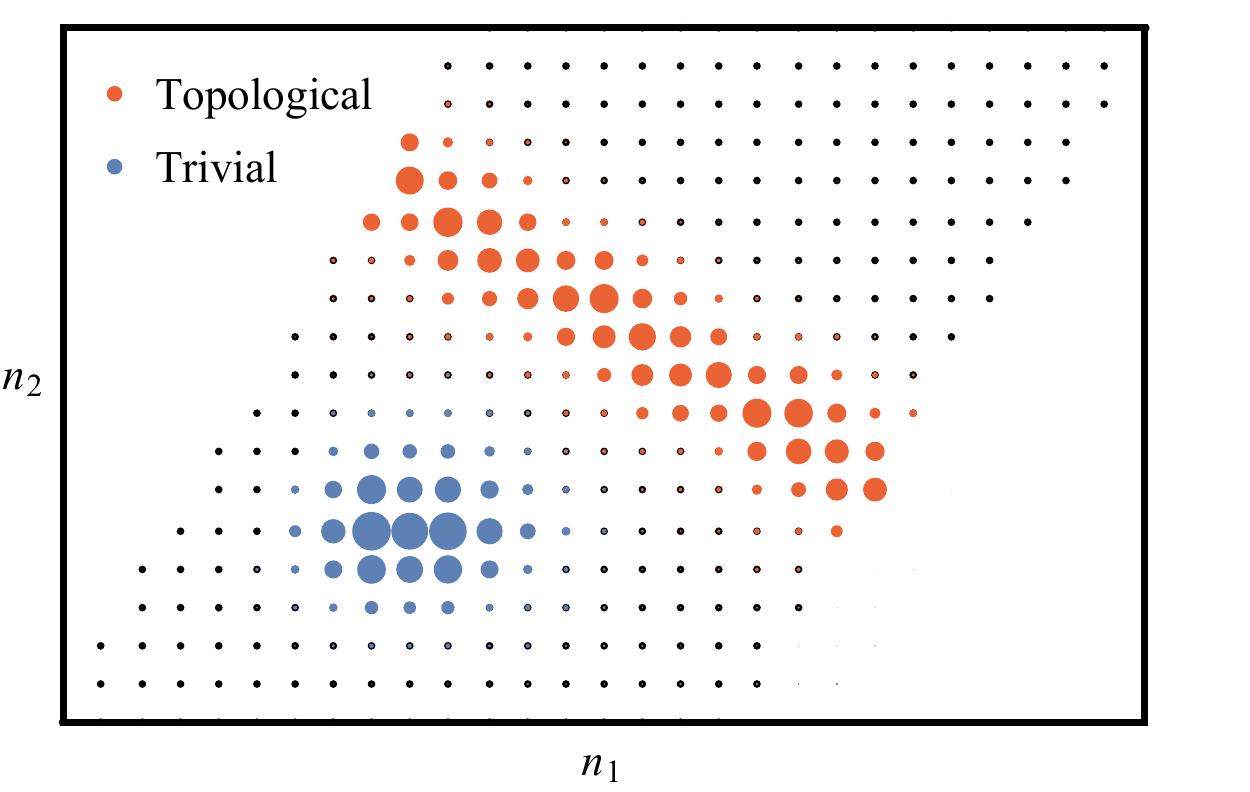}
\caption{
\emph{Localisation and delocalisation on the frequency lattice:} The support of the quasi-energy states on the frequency lattice in the commensurate approximation (pink region in Fig.~\ref{Fig:FreqLattice}). Each red/blue disk is centred on a lattice site $\Bn$ and has an area $\propto \log \langle{\tilde\phi^j_\Bn(\Bt_0)}\ket{\tilde\phi^j_\Bn(\Bt_0)}$. The topological states (red) are delocalised and encircle the cylinder, whereas the trivial states (blue) are localised. Data from $H_\mathrm{CD}$~\eqref{eq:topomodel} with $(p,q)=(8,5)$, and $m=1$ (left) and $m=3$ (right).
}
\label{Fig:QESSupport}
\end{center}
\end{figure}

\begin{figure}
\begin{center}
\includegraphics[width=\columnwidth]{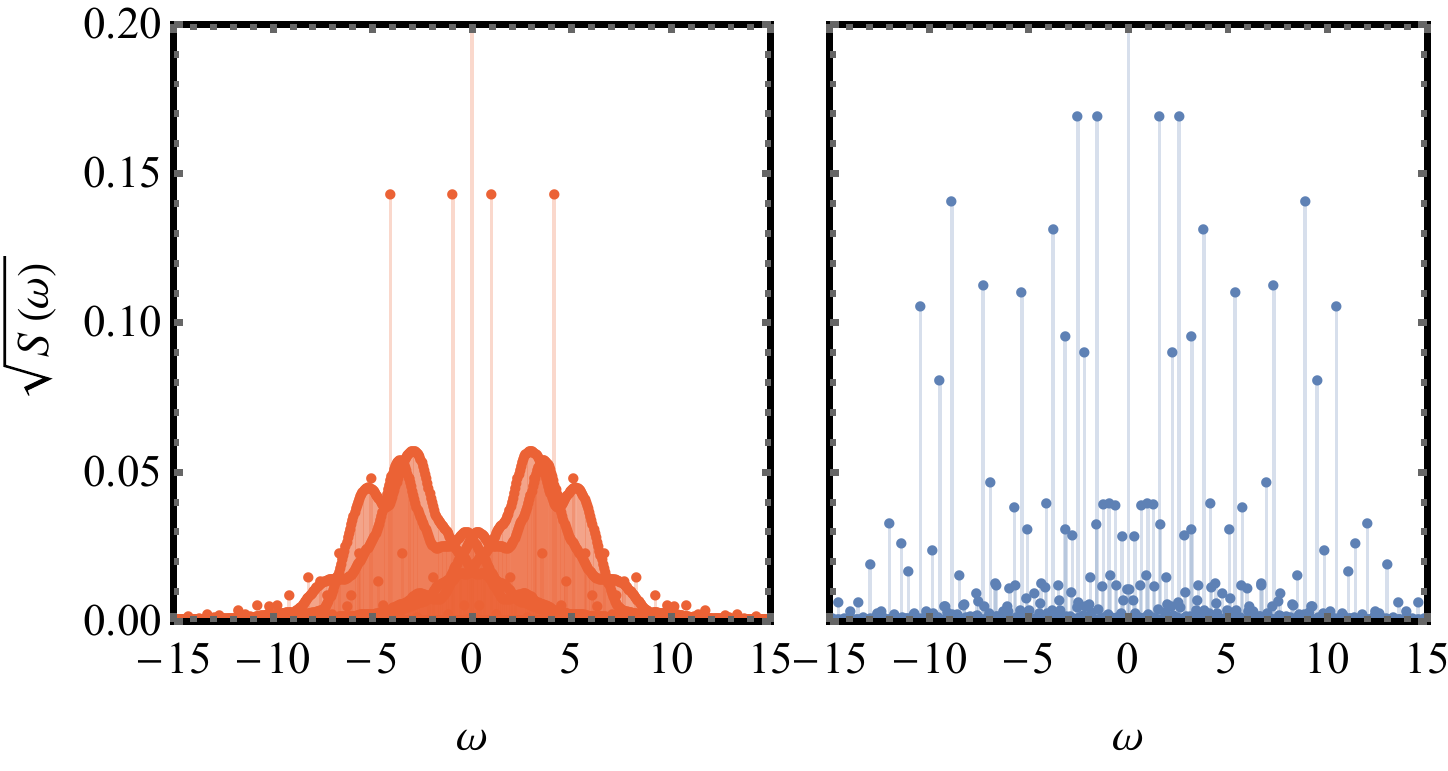}
\caption{
\emph{Spectral properties of expectation values:} Square root of the mean power spectrum for topological (left) and trivial (right) classes of dynamics in the commensurate approximation, showing a nascent region of dense spectrum only in the topological case. The power spectrum is averaged over initial states and polarisation axes $\vec{a}$ of operators $A=\vec{a}\cdot\vec{\sigma}$ for $|\vec{a}|=1$. Data from $H_\mathrm{CD}$~\eqref{eq:topomodel} with $(p,q)=(89,55)$, and $m=1$ (red) and $m=3$ (blue).
}
\label{Fig:OpSpec_fig}
\end{center}
\end{figure}

A quasi-energy state $\ket{\tilde\phi^j(\Bt_0)}$ belonging to a band with $C_j \neq 0$ is delocalised on the frequency lattice in the direction perpendicular to the electric field $\vec{\Omega}$.
Indeed, in order for the state to be sensitive to flux threading through the cylinder or to pump energy indefinitely, it \emph{has} to be delocalised. See Fig.~\ref{Fig:QESSupport}.

As $\ket{\tilde\phi^j_\Bn(\Bt_0)}$ are the Fourier components of $\ket{\phi^j(\Bt_t)}$ with frequency $\omega = \epsilon_j +  \Bn\cdot\Bw$, the state $\ket{\phi^j(\Bt_t)}$ has a dense Fourier spectrum for $C_j \neq 0$. 
Expectation values are therefore aperiodic in the topological class.
If $C_j=0$, then the quasi-energy states are localized on the frequency lattice and the state $\ket{\phi^j(\Bt_t)}$ has a sparse Fourier spectrum that can be approximated to any desired accuracy with a finite number of components. Expectation values are quasi-periodic in time in this case. See Fig.~\ref{Fig:OpSpec_fig}.

\section{Stability of the topological class}
\label{sec:stability}

The topological class does not extend to a phase because of need of an exact level crossings in the quasi-energy band structure.
Recall that the Chern numbers satisfy the sum rule $\sum_j C_j = 0$.
If there is a band $j$ with $C_j \neq 0$ in the spectrum, then there must be another band $j'$ with a Chern number of the opposite sign by the sum rule.
As the Chern number sets the gradient of the dispersion, the bands $j$ and $j'$ must cross.
These crossings are visible in Fig.~\ref{Fig:BandStructPlot}. 
The topological class of dynamics is thus realised only if the quasi-energy operator $K(\Bt_0)$ has exact degeneracies at some $\Bt_0$.
We expect that the exact degeneracy splits on perturbing $K(\Bt_0)$. 
Thus, the topological case is finely tuned, and only the trivial class with all $C_j=0$ is stable to perturbation.

Despite this generic instability, in this section we study two constructions which realise the topological dynamics in settings amenable to experiment. We start from a model (introduced in Ref.~\cite{martin2017topological}) which realises the topological phase exactly in the adiabatic limit. Firstly, we show that at finite drive rate this immediately yields a long pre-thermal period for which the dynamics of the topological class is observed; secondly, we use a counter-diabatic correction to produce an explicit, finely tuned model, which realises the topological class of dynamics indefinitely, at any finite drive rate, and which is exponentially dominated by a finite bandwidth of drive frequencies.

We first consider the Chern insulator (CI) model
\begin{equation}
H_\mathrm{CI}(\Bt_t) = \begin{pmatrix}
\sin \theta_{t1}  \\
\sin \theta_{t2}  \\
m- \cos \theta_{t1} - \cos \theta_{t2} \\
\end{pmatrix}\cdot \vec{\sigma}
\label{eq:H_BHZ}
\end{equation}
previously introduced in~\eqref{eq:Hintro}, here $\vec{\sigma} = (\x,\y,\z)$, and $\Bt_t = \Bt_0 + \Bw t$.
We are motivated to study this model by the analogy to Hall physics (see Sec.~\ref{sec:pumppump}): Eq.~\eqref{eq:H_BHZ} is a well-known Chern insulator~\cite{qi2006topological,bernevig2006quantum} where we have made the replacement $(k_x,k_y) \to (\theta_{t1},\theta_{t2})$~\footnote{Note that the quasi-energy band-structure described in Sec.~\ref{sec:QETopo} is different from the usual band-structure of the model~\eqref{eq:H_BHZ} as we have included the non-perturbative effects of the field $\Bw$ as opposed to the usual Kubo-formula calculation where the effect of the electric field is accounted for perturbatively.}. It follows that for, $0\leq|m|<2$ the \emph{instantaneous eigenstates} of $H_\mathrm{CI}$ form bands with non-trivial Chern numbers: $(C_1,C_2)=(1,-1)$ for $0<m<2$ which switch signs to $(C_1,C_2)=(-1,1)$ for $-2<m<0$.

In the precise limit $\Omega_1,\Omega_2 \to 0$ it follows from the adiabatic theorem that the quasi-energy-states are given by the instantaneous eigenstates of $H_\mathrm{CI}$, and thus inherit the non trivial Chern numbers of the Hall problem. These non-trivial Chern numbers constitute a realisation of the topological class of dynamics. 

\subsection{Pre-thermal topological dynamics}
\label{sec:notopoBHZ}

\begin{figure}
\begin{center}
\includegraphics[width=\columnwidth]{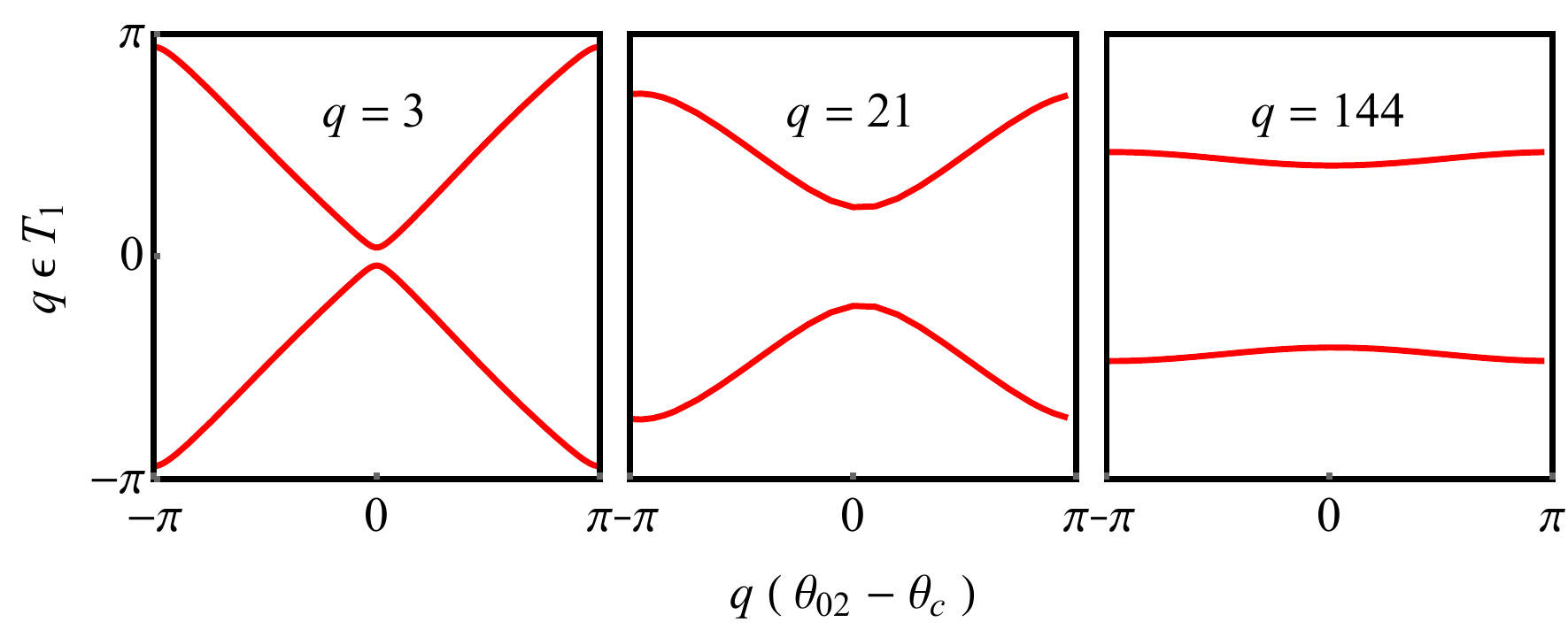}
\caption{
\emph{Quasi-energy band-structure of $H_\mathrm{CI}$~\eqref{eq:H_BHZ} as the quasi-periodic limit is approached:} 
Plots of quasi-energy $\epsilon_j$ versus the initial phase $\theta_{02} - \theta_c$, where $\theta_{02} = \theta_c$ is the point of minimum gap. 
As $q$ increases toward the incommensurate limit, the quasi-energy bands flatten. Note the rescaling by $q$ vs. Fig~\ref{Fig:BandStructPlot}.
Parameters: $\Omega_1 = 2 \pi/15$, $m=1$.
}
\label{Fig:BS}
\end{center}
\end{figure}

At finite drive frequencies $\Omega_1$ and $\Omega_2$, the dynamical states of the CI model fail to follow the adiabatic eigenstates, and the system heats by Landau-Zener excitation. 

In the low frequency limit, the Landau-Zener rate of excitation $1/\tau$ is exponentially small in the rate of change of the Hamiltonian~\cite{landau1932theorie,zener1932non,stueckelberg1933theorie,majorana1932atomi}
\begin{equation}
\log\tau \sim |\Bw|^{-1} \sim T_1, T_2.
\label{eq:LZtime}
\end{equation}
In the pre-thermal regime, $0 < t \ll \tau$, this rate is negligible and the deviation from the adiabatic limit is small. 
A qudit prepared in an instantaneous eigenstate remains close to one, and the dynamics are controlled by the topological class of the strict adiabatic limit.
This pre-thermal regime is exponentially long in the drive period $T_1,T_2$, making the regime accessible to experiment.

In the commensurate case the total period, $T = q T_1 = p T_2$, provides an additional time scale. 
In the adiabatic limit the Floquet states of the commensurate problem also have non-zero quasi-energy gradient (see~\eqref{eq:WC} and Fig.~\ref{Fig:BandStructPlot}), and so exhibit the properties of the topological class of dynamics. At finite drive rate, if the period $T \ll \tau$ then the effect of Landau-Zener excitation within a period is small, the Floquet states are only weakly perturbed, the quasi-energies are close to the quantised values, and the system continues to exhibit the topological dynamics, though the average pumping is no longer quantised. The topological dynamics are exhibited for generic initial conditions except for an exponentially small set of initial conditions close to the avoided crossing of quasi-energy bands. For initial conditions close to the avoided crossing the Floquet states are strongly altered, and a state prepared in an instantaneous eigenstate will scatter into other eigenstates on the timescale $\tau$. 

The band-structure of the commensurate system is depicted in Fig.~\ref{Fig:BS}. 
For $q T_1\ll \tau$ (left panel) the avoided crossing is small, and for much of the Floquet zone the quasi-energy gradient is close to the quantised value~\eqref{eq:grad_eps}, from this the dynamical properties of the topological class of dynamics follow. As $q$ is increased, lengthening the period, and bringing the system closer to the incommensurate limit, the avoided crossing begins to dominate the band-structure, and the quasi-energy levels approach their flat (topologically trivial) limiting form. For $q T_1 \gg \tau$ the signatures of the topological dynamics are lost on the shorter timescale $\tau$.

\subsubsection{Energy pumping in the pre-thermal regime}

\begin{figure}
\begin{center}
\includegraphics[width=0.95\columnwidth]{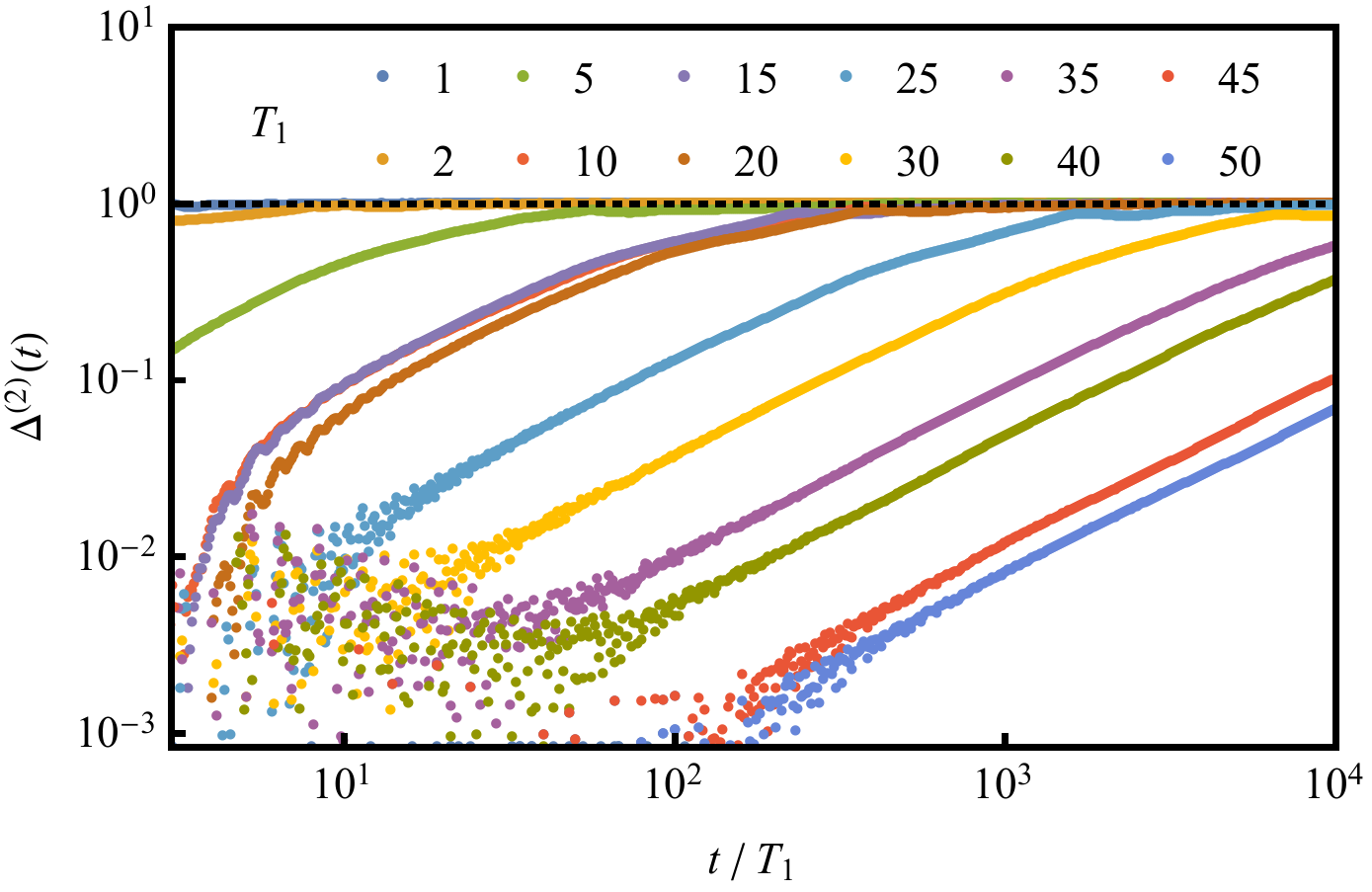}
\includegraphics[width=0.95\columnwidth]{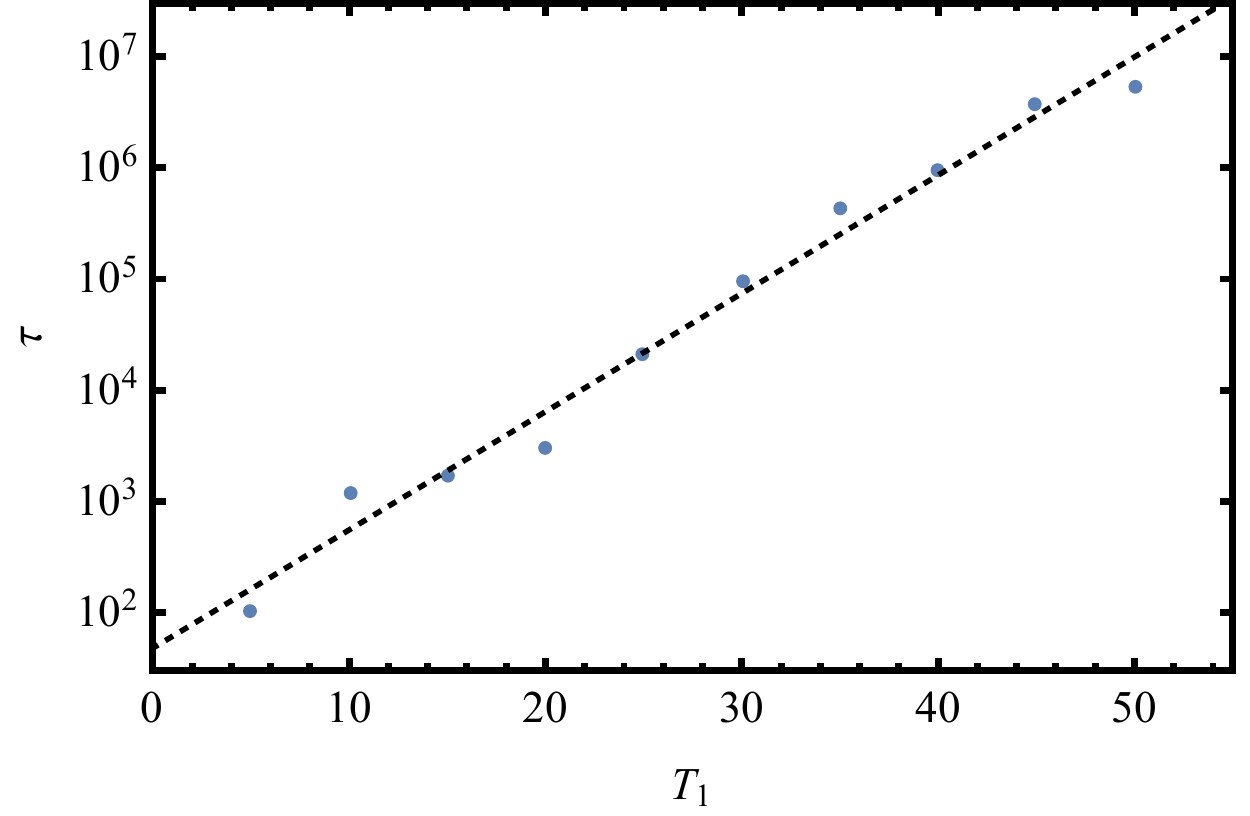}
\caption{
\emph{Decay of quantised pumping in $H_\mathrm{CI}$~\eqref{eq:H_BHZ}}: (Upper panel) The normalised deviation $\Delta^{(2)}(t)$ (see Eq.~\eqref{eq:normdev}) of the time averaged pump power, maximised over initial states $\ket{\psi_0}$, from the quantised value. At $T_1<\infty$ (i.e. $\Omega_1,\Omega_2>0$) the pumping is found to decay at an initially linear rate which we estimate by fit to each series over the range $10^{-2}< \Delta^{(2)}(t) < 0.4$. (Lower panel) This decay rate is exponentially small in the drive rate. Each data point in the upper plot is averaged over $N=4000$ trajectories with random $\Bt_0$. Data for model $H_\mathrm{CI}$~\eqref{eq:H_BHZ} with $m=1$ and $\Bt_0$ drawn uniformly from the Floquet zone.
}
\label{Fig:Pumping}
\end{center}
\end{figure}

The Landau-Zener scaling of the loss of the dynamical signatures of the topological class is confirmed by analysing the energy pumped by the system. 

In the Heisenberg picture the instantaneous power of the 2nd drive is given by the operator (see~\eqref{eq:P2},~\eqref{eq:deltaE})
\begin{equation}
P^{(2)}(t) = \Omega_2 U^\dag(t,0;\Bt_0) \partial_{\theta_{02}} H(\Bt_t) U(t,0;\Bt_0).
\end{equation}
The mean power over an interval $[0,t]$ maximised over initial states is then given by
\begin{equation}
P_{\max}^{(2)}(t) = \max_{\ket{\psi_0}}\qexp{\left[\frac{1}{t}\int_0^t \d t' P^{(2)}(t')\right]}{\psi_0}.
\end{equation}
By using this measure we avoid the question of which initial states exhibit pumping most clearly over finite times.

We compare this with the theoretical value for a topological model with Chern number $C=1$ of $P = \Omega_1\Omega_2/2\pi$ given by~\eqref{eq:pumppower}. Deviation from the topological value is captured by the normalised deviation of the pump power $\bar{P}_j^{(n)}$ from the theoretically maximal value $P$
\begin{equation}
\Delta^{(n)}(t) = 1-\frac{P_{\max}^{(n)}}{P}.
\label{eq:normdev}
\end{equation}
In the upper panel of Fig.~\ref{Fig:Pumping} we plot $\Delta^{(n)}(t)$ vs $\Omega_1 t$ for various values of $T_1=2\pi/\Omega_1$ with fixed $\Omega_2/\Omega_1 = (1+\sqrt{5})/2$. The eventual decay of pumping to zero results in the system converging to asymptotic value $\Delta^{(n)} = 1$ in all cases.

For small times one finds the decay is linear, 
\begin{equation}
\Delta^{(n)}(t) = t/\tau + O(t^2/\tau^2).
\end{equation}
The values of $\tau$ are extracted by a linear fit to the data from the upper panel of Fig.~\ref{Fig:Pumping} in the region $\Delta^{(n)}<0.6$, i.e. before the curves begin to flatten into their asymptotic values $\Delta^{(n)} = 1$. These extracted values are plotted versus $T_1$ in the lower panel. We see that the  decay time $\tau$ is exponentially long in the inverse drive rate, 
\begin{equation}
\log\tau \sim T_1,T_2
\label{eq:tLZ}
\end{equation}
consistent with Landau-Zener excitation.

\subsection{Finite rate counter diabatic driving}
\label{sec:CD}

\begin{figure}
\begin{center}
\includegraphics[width=\columnwidth]{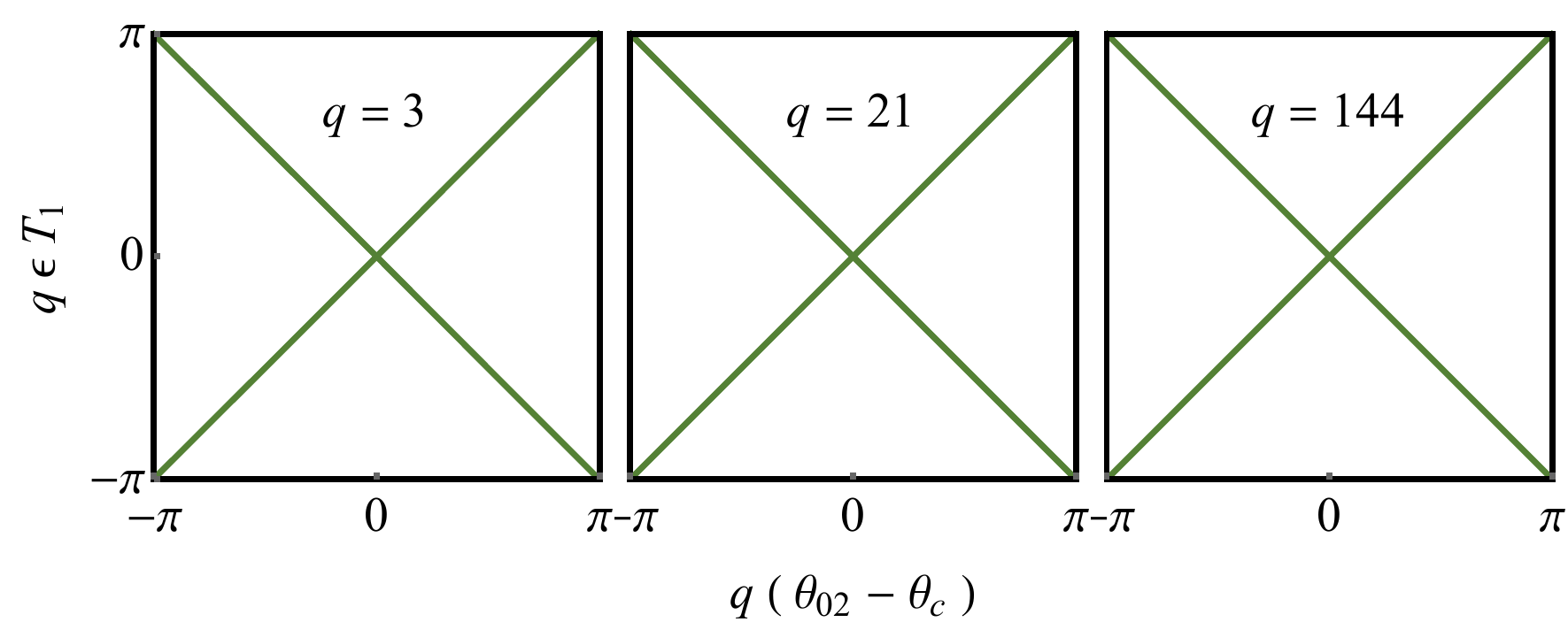}
\caption{
\emph{Quasi-energy band-structure of $H_{\mathrm{CD}}$~\eqref{eq:topomodel} as a function of $q$:} The quasi-energy $\epsilon_j$ versus the initial phase $\theta_{02} - \theta_c$ where $\theta_{02} = \theta_c$ is the point of minimum gap in the extended zone scheme. The counter-diabatic term in~\eqref{Eq:VForm} protects the linearly dispersing bands and exact level crossings at all $q$. The dynamical class is topological as $q \to \infty$. Parameters $\Omega_1 = 2 \pi/15$, $m=1$ as in Fig.~\ref{Fig:BS}.
}
\label{Fig:BS2}
\end{center}
\end{figure}

\begin{figure}
\begin{center}
\includegraphics[width=\columnwidth]{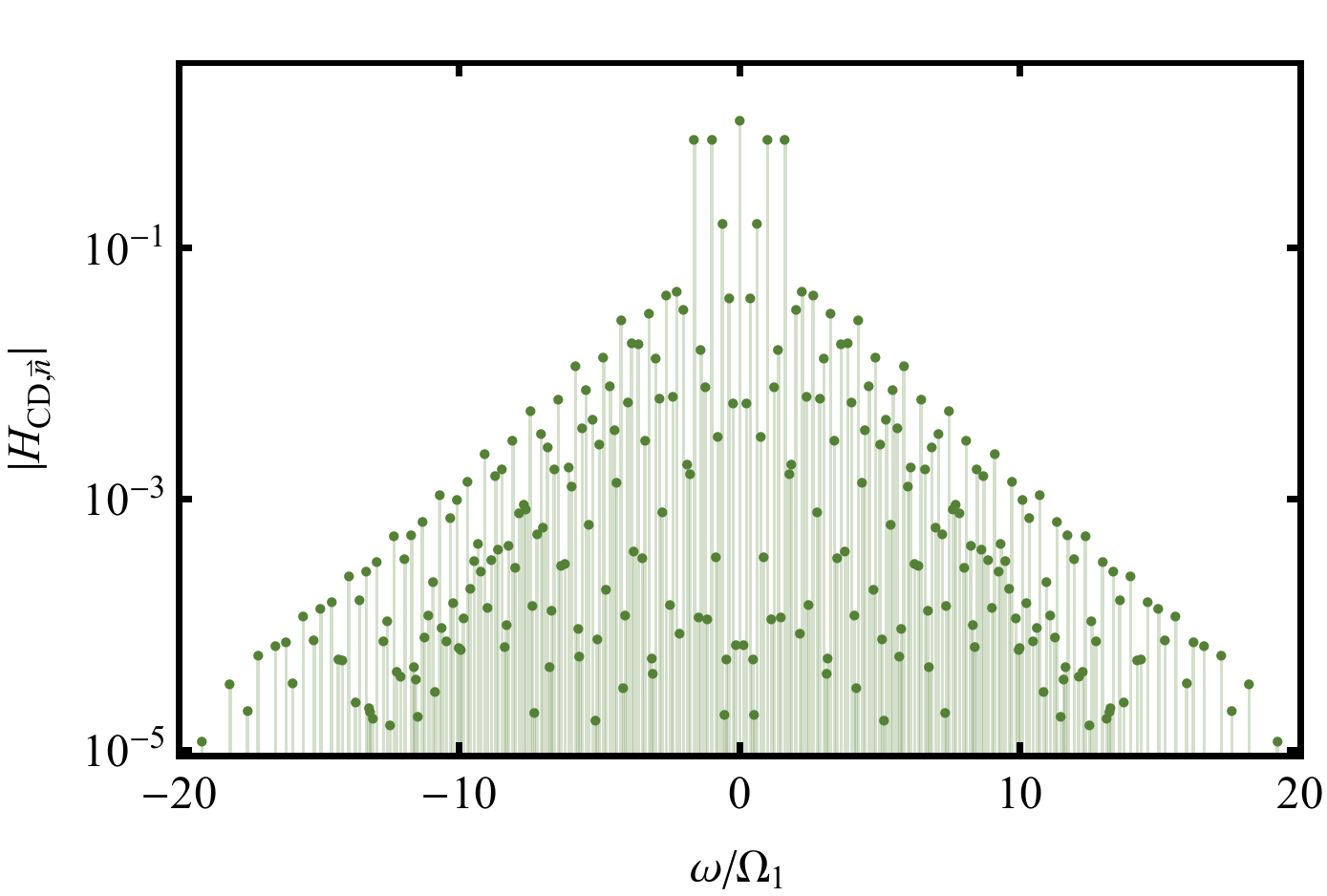}
\caption{
\emph{Fourier spectrum of $H_\mathrm{CD}$~\eqref{eq:topomodel}:} The Frobenius norm of the largest 265 Fourier components $H_{\mathrm{CD},\Bn}$ of the Hamiltonian $H_{\mathrm{CD}}$  (defined via $H_\mathrm{CD}(\Bt_t) = \sum_\Bn H_{\mathrm{CD},\Bn} \mathrm{e}^{- \i \Bn \cdot \Bt_t}$) for $m=1$. These are plotted versus their corresponding frequency $\omega= \Bn\cdot\Bw$. The spectrum is exponentially decaying away from $\omega = 0$ due to the analyticity of $H_{\mathrm{CD}}$.
}
\label{Fig:HFourier}
\end{center}
\end{figure}

Adding a counter-diabatic correction term to the Hamiltonian prevents the Landau-Zener processes that destroy the dynamical signatures of the topological class on time-scales $t \gtrsim \tau$~\cite{del2013shortcuts,sels2017minimizing}. 
Using this method we obtain, to our knowledge, the first analytic Hamiltonian which realises topological dynamics in a quasi-periodically driven system indefinitely. This model has finite frequency bandwidth and finite drive rate, making it amenable to experimental study.

The counter-diabatic correction $V$ precisely cancels the matrix elements coupling the instantaneous eigenstates. For any time-dependent Hamiltonian $H(t)$, the condition for the cancellation is:
\begin{equation}
\com{\i \partial_t H + \com{H}{V}}{H} = 0.
\label{eq:CD}
\end{equation}
For a spin-$1/2$ traceless Hamiltonian~\eqref{eq:CD} has the solution
\begin{equation}
V = \frac{\i}{2}\frac{\com{\partial_t H}{H}}{ \tr{H^2}} + u H + v \bm{1}
\label{Eq:VForm}
\end{equation}
for free parameters $u,v$. Without loss of generality, we take $u=v=0$. 

The quasi-energy-states of the corrected model
\begin{equation}
H_\mathrm{CD} = H_\mathrm{CI}+V
\label{eq:topomodel}
\end{equation}
are the instantaneous eigenstates of $H_\mathrm{CI}$~\eqref{eq:H_BHZ}. Thus, if $H$ is in the topological class in the strictly adiabatic limit, then $H_\mathrm{CD}$ is in the topological class for any drive frequency. The resulting topological quasi-energy band-structure is verified in Fig.~\ref{Fig:BS2} (using the same parameters as in Fig.~\ref{Fig:BS}).

The norm of the Fourier amplitudes of $H_{\mathrm{CD}}$ for the Chern insulator model is shown in Fig.~\ref{Fig:HFourier}.
We see that the norm decays exponentially away from zero frequency.
Thus, the corresponding frequency lattice model has exponentially decaying hopping terms.
Approximating $H_\mathrm{CD}$ by truncating to the $N$ largest Fourier amplitudes leads to an exponentially small in $N$ error term in $H_{\mathrm{CD}}(t)$.
This truncation leads to hybridization of the instantaneous eigenstates, as in the previous section.
The dynamics of the topological class are then lost after an exponentially long pre-thermal regime $t \ll \tau$ with $\log \tau \sim N$.

\subsubsection{Numerical observation of the topological class with $H_\mathrm{CD}$ in the Chern insulator model}

We numerically verify that $H_{\mathrm{CD}}(t) = H_{\mathrm{CI}}(t) + V(t)$ realizes the topological class of dynamics for $0<|m|<2$ and the trivial class of dynamics for $|m|>2$. 

\paragraph{Stable topological band-structure:} Fig.~\ref{Fig:BandStructPlot} shows the quasi-energy band structure of $H_{\mathrm{CD}}(t)$ at $q=5$. Without the counter-diabatic correction, the bands flatten with increasing $q$ for any $m$, as shown in Fig.~\ref{Fig:BS}. $V(t)$ protects the exact level crossings for $0<|m|<2$ as $q \to \infty$ in Fig.~\ref{Fig:BS2}, so that the dynamical class is topological in the incommensurate limit. 

\paragraph{Net energy pumping:}

The total energy pumped as a function of time is shown in Fig.~\ref{Fig:PumpingPlot} for the topological (red) and trivial (blue) classes. Asymptotically, both curves follow the theoretical prediction in~\eqref{eq:pumppower}.

\paragraph{Divergence of trajectories:}

Fig.~\ref{Fig:PumpingPlot} shows the dynamics of the Bures Angle~\eqref{eq:BuresAngle} for nearby initial states for topological (red) and trivial (blue) dynamics. Asymptotically, both curves follow the theoretical prediction in~\eqref{eq:errorgrowth}.
\paragraph{Delocalisation on the frequency lattice:}

In the topological class, the quasi-energy states are delocalised on the frequency lattice (Sec.~\ref{sec:deloc}). Fig.~\ref{Fig:QESSupport} qualitatively shows this. In App.~\ref{app:Scaling}, we show quantitatively that the scaling in the commensurate limit is consistent with frequency lattice quasi-energy states that are delocalised in the topological class and localised in the trivial class.

\paragraph{Dense Fourier spectra of observables:}

The Fourier amplitude $A(\omega)$ of an expectation value $\cexp{A(t)}$ is given by
\begin{equation}
\begin{aligned}
\qexp{A}{\psi(t)} &= \sum_{j,j'}\sum_{\Bn,\Bm} \alpha_j^*\alpha_{j'} \bra{\phi_{\Bm}^j} A \ket{\phi_{\Bn+\Bm}^{j'}}\mathrm{e}^{-\i(\Bw\cdot\Bn+\epsilon_{j'}-\epsilon_j)t}\\
& =\sum_\omega A(\omega) \mathrm{e}^{- \i \omega t},
\end{aligned}
\label{eq:FTobs}
\end{equation}
for some observable $A$ and a generic initial state~\eqref{eq:genstate}.
We characterise the Fourier amplitude by the mean power spectrum
\begin{equation}
\begin{aligned}
S(\omega) &= \left[ |A(\omega)|^2 \right]_{A,\ket{\psi_0}} 
\end{aligned}
\label{eq:PSobs}
\end{equation} 
where the right-hand side is averaged over operators of the form $A = \vec{a}\cdot\vec{\sigma}$ with $\vec{a}$ drawn uniformly from the unit sphere, and initial states drawn uniformly from the Bloch sphere. For details of this calculation see App.~\ref{app:Sw}.

$\sqrt{S(\omega)}$ is the root-mean-square magnitude of the Fourier coefficients of $\qexp{A}{\psi(t)}$. The values of $\sqrt{S(\omega)}$ are plotted as lollipops for the trivial and topological cases in Fig.~\ref{Fig:OpSpec_fig} using the commensurate approximation. Spectra for the topological and trivial cases are found to have a pure-point part, whereas only the topological case has a continuous part. The nascent continuous part of the topological spectrum is visible in Fig.~\ref{Fig:OpSpec_fig} where the lollipops appear to blur together into lines~\footnote{We note that the topological power spectra is seen here to be comprised of three lines, corresponding to the frequencies $\Bw \cdot \Bn$ and $\Bw \cdot \Bn \pm \omega_{jj'}$ where $\omega_{jj'} = \epsilon_j - \epsilon_{j'}$ is the difference between consecutive quasi-energies. In the quasi-periodic limit the separation between the points making up these curves tends to zero, and they cannot be resolved.}.

We further verify in App.~\ref{app:Scaling} via scaling in the commensurate approximation that in the quasi-periodic limit, the spectrum $A(\omega)$ becomes dense in the topological class and remains sparse in the trivial class.

\section{Summary and discussion}
\label{sec:discussion}

We have classified the dynamical properties of a $d$-level qu$d$it driven by two tones with incommensurate frequencies using $d$ integer Chern numbers. 
The generalization of Floquet theory to the two-tone setting identifies a two-dimensional tight-binding model in frequency (Fourier) space, the Hamiltonian of this model is the quasi-energy operator $K$.
We organized the eigenstates of $K$ with distinct time dependences into a quasi-energy band structure on the torus of initial drive phases, with integer Chern number $C_j$ associated to each band $j$. 

Starting from a generic initial qudit state, we observe (1) the pumping of energy between the drives, (2) sensitivity to the drive phases at $t=0$ and (3) aperiodic dynamics of observables, {\em only} in the topological class (with at least one $C_j \neq 0$). 
In contrast, the phenomenology of the trivial class (with all $C_j=0$) is the same as the one-tone driven case.

Although the topological class does not extend to a phase with a finite volume in the parameter space, it leads to an exponentially long pre-thermal regime in the near-adiabatic limit. For finite drive frequencies (non-adiabatic regime), we constructed (fine-tuned) models that belong to the topological class using counter-diabatic methods. These correspond to introducing an infinite number of extra hopping elements on the frequency lattice, with magnitude decaying exponentially with the hopping distance.

More generally, the band crossing required to realize the topological class can be accomplished by introducing extra tuning parameters in the Hamiltonian. In the case of counter-diabatic driving the extra parameters are the higher harmonics (which correspond to long range hops on the frequency lattice). Based on the general theory~\cite{von1929behaviour}, two band touching requires tuning of three parameters for a general unitary evolution operators or fewer in the presence of extra symmetry. The number of needed tuning parameters defines the codimension of the band touching manifold in the full parameter space. Note that if the codimension is greater than one (e.g. point in 2D, or a line in 3D, both corresponding to codimension two), the manifold cannot split the parameter space into disjoint subspaces, and the band-touching points cannot demarcate the boundaries of distinct trivial phases.

One may also access additional tuning parameters by considering more than two incommensurate drives, $n_\mathrm{f}>2$. In this case, there are $n_\mathrm{f}-1$ independent relative phases, and the quasi-energy $\epsilon_j(\Bt_0)$ is a non-trivial function of $n_\mathrm{f}-1$ variables, facilitating the quasi-energy level crossings. The physical significance of such models and possible manifestations of topological phases in this case (analogous to frequency pumping for $n_\mathrm{f} = 2$) are open and interesting questions.

Furthermore, in future research it will be interesting to investigate how extra symmetries and extra parameters can be used to access the topological classes. For instance, an analogue of time reversal symmetry at special points in the Floquet zone can lead to Kramers doublets in the quasi-energy spectrum, and therefore exact level crossings in the quasi-energy band structure.
The role of symmetry and its effects on the dynamical classification can be investigated in driven qudit models inspired by the momentum space representation of the Kane-Mele model~\cite{kane2005quantum} or models with a Zak phase~\cite{zak1989berry,delplace2011zak}.

Dissipative forces provide another route to stabilize the topological class, as noted in Ref.~\cite{martin2017topological}.
If the relaxation time of the qudit is much smaller than the Landau-Zener mixing time~\eqref{eq:LZtime}, then the qudit remains close to the instantaneous ground state indefinitely.
The qudit will therefore pump energy between the drives at a nearly quantized rate even at finite drive frequencies if the quasi-energy band associated with the instantaneous ground state has a non-zero Chern number.
However, such inherently quantum effects as the sensitivity to initial drive phases which rely on phase coherence will be lost.
Understanding the effects of different types of dissipative forces on the properties of the topological class is crucial for the experimental observation of the topological class. 

Finally, the combination of spatial and synthetic dimensions may result in new dynamical classes with no equilibrium counterpart. \citet{rudner2013anomalous} followed by Refs.~\cite{nathan2015topological,roy2016abelian,else2016classification,potter2016classification,roy2017periodic} developed classifications for the Floquet unitaries of extended systems and showed new topological orders of robust edge modes in periodically driven trivial insulators.

It would be very interesting to extend that framework for incommensurately driven insulators.

\begin{acknowledgements}
We are grateful to C. Baldwin, E. Berg, J. Chalker, S. Gopalakrishnan, S. Kourtis, C. Laumann, P. Mehta, F. Nathan, A. Polkovnikov, G. Refael, and D. Sels for many useful discussions.

We thank the Kavli Institute for Theoretical Physics (KITP) in Santa Barbara for their hospitality during the early stages of this work, the National Science Foundation (NSF) under Grant No. NSF PHY-1748958 for supporting KITP, and the BU shared computing cluster for computational facilities.

AC acknowledges support from the NSF through grant No. DMR-1752759.
IM was supported by the Department of Energy, Office of Science, Materials Science and Engineering Division. 

\end{acknowledgements}

\bibliography{QPDriving_bib}

\appendix
\begin{widetext}
\section{Derivation of Eq.~\eqref{eq:dedtheta}}
\label{app:dedtheta}
In this appendix we show a derivation of the result
\begin{align}
\bra{\tilde\phi^j(\Bt_0)} \partial_{\theta_{02}} K(\Bt_0) \ket{\tilde\phi^j(\Bt_0)} = \lim_{t \to \infty} \frac{1}{t} \int_0^{t} \d s \bra{\phi^j(\Bt_s)} \partial_{\theta_{02}} H(\Bt_s)\ket{\phi^j(\Bt_s)}.
\end{align}
First using the relation~\eqref{eq:gauge_fix_phi} we can write 
\begin{equation}
\bra{\phi^j(\Bt_s)} \partial_{\theta_{02}} H(\Bt_s)\ket{\phi^j(\Bt_s)} = \bra{\phi^j(s;\Bt_0)} \partial_{\theta_{02}} H(\Bt_s)\ket{\phi^j(s;\Bt_0)}.
\end{equation}
Then substituting for the Fourier representations we find
\begin{equation}
\begin{aligned}
\bra{\phi^j(s;\Bt_0)} \partial_{\theta_{02}} H(\Bt_s)\ket{\phi^j(s;\Bt_0)} = \sum_{\Bn,\Bm,\Bk}  \mathrm{e}^{\i (\Bn - \Bm - \Bk)\cdot \Bw s}\bra{\tilde\phi^j_\Bn(\Bt_0)} (-\i k_2)H_{\Bk} \e^{- \i \Bk \cdot\Bt_0  }\ket{\tilde\phi^j_\Bm(\Bt_0)}.
\end{aligned}
\end{equation}
The time integral then selects the terms $\Bn-\Bm=\Bk$ and we find
\begin{equation}
\lim_{t \to \infty} \frac{1}{t} \int_0^{t} \d s \bra{\phi^j(\Bt_s)} \partial_{\theta_{02}} H(\Bt_s)\ket{\phi^j(\Bt_s)} = \sum_{\Bn,\Bm} \bra{\tilde\phi^j_\Bn(\Bt_0)} (-\i (n_2-m_2))H_{\Bn-\Bm} \e^{- \i (\Bn - \Bm) \cdot\Bt_0  }\ket{\tilde\phi^j_\Bm(\Bt_0)}.
\end{equation}
Recognising the term in the middle from~\eqref{eq:K} it is easily shown that
\begin{equation}
\partial_{\theta_{02}}K =\sum_{\Bn \Bm} \left[ (-\i (n_2-m_2)) H_{\Bn-\Bm} \e^{-\i (\Bn - \Bm )\cdot \Bt_0} \right] \otimes \ket{\Bn} \bra{\Bm}.
\end{equation}
By substitution we then see that
\begin{equation}
\begin{aligned}
\lim_{t \to \infty} \frac{1}{t} \int_0^{t} \d s \bra{\phi^j(\Bt_s)} \partial_{\theta_{02}} H(\Bt_s)\ket{\phi^j(\Bt_s)} &= \left(\sum_\Bn \bra{\tilde\phi^j_\Bn(\Bt_0)} \otimes \bra{\Bn}\right)\partial_{\theta_{02}}K \left(\sum_\Bm \ket{\tilde\phi^j_\Bm(\Bt_0)} \otimes \ket{\Bm}\right) \\
& = \bra{\tilde\phi^j(\Bt_0)} \partial_{\theta_{02}} K(\Bt_0) \ket{\tilde\phi^j(\Bt_0)}
\end{aligned}
\end{equation}
which shows the proposition.

\section{$C_j=0$ implies monodromy}
\label{app:Monodromy}

In this appendix we show that (i) given a quasi-energy state $\ket{\phi^j_\mathrm{S}(0,\theta_2)}$, defined over the line $\theta_{1}=0$, in  smooth gauge; (ii) if $C_j =0$, then it is possible to construct a gauge in which the monodromy relation~\eqref{eq:monodromy} is satisfied.

It is always possible to define a smooth gauge $\ket{\phi^j_\mathrm{S}(0,\theta_2)}$ over the line $\theta_1=0$. For example, given any gauge $\ket{\phi^j(0,\theta_2)}$ a smooth gauge may be constructed by multiplying by a $\theta_2$ dependent phase such that the phase the projection onto an arbitrary reference state $\ket{r}$ to be real $\braket{r}{\phi^j_\mathrm{S}(0,\theta_2)} \in \mathbb{R}$. For a smooth gauge we can use a construction analogous to~\eqref{eq:extendmono} to define
\begin{equation}
\ket{\phi^j_\mathrm{S}(\Omega_1 t , \Omega_2 t + \theta_2)} = \e^{\i \epsilon_j t}U(t,0;0,\theta_2)\ket{\phi^j_\mathrm{S}(0,\theta_2)}
\label{eq:extendmono2}
\end{equation}
where we have used that for $C_j=0$ that $\epsilon_j$ is independent of $\Bt$. The gauge is smooth in interior of the Floquet zone by construction. The gauge is then globally smooth if $\ket{\phi^j_\mathrm{S}(2 \pi,\theta_2)} = \ket{\phi^j_\mathrm{S}(0,\theta_2)}$. However in general there is a phase difference
\begin{equation}
\begin{aligned}
z(\theta_2) &= \braket{\phi^j_\mathrm{S}(0,\theta_2+2\pi \beta)}{\phi^j_\mathrm{S}(2 \pi,\theta_2+2\pi \beta)} \\
& = \bra{\phi^j_\mathrm{S}(0,\theta_2+2 \pi \beta)} U(T_1,0;0,\theta_2)\ket{\phi^j_\mathrm{S}(0,\theta_2)}.
\end{aligned}
\end{equation}
For for $C_j=0$ the quasi-energy $\epsilon_j$ is independent of $\theta_2$ and this phase is given by
\begin{equation}
z(\theta_2) = \e^{-\i \epsilon_j T_1} u^*(\theta_2+2 \pi \beta)u(\theta_2)
\label{eq:zform}
\end{equation}
where $u(\theta_2) = \braket{\phi^j(0,\theta_2)}{\phi^j_\mathrm{S}(0,\theta_2)}$ is the phase difference between the smooth gauge $\ket{\phi^j_\mathrm{S}(2 \pi,\theta_2)}$ and $\ket{\phi^j(0,\theta_2)}$, the gauge of Sec.~\ref{sec:guagefix}. The phase $z(\theta_2)$ is smooth by construction, and for $C_j=0$ has no winding number
\begin{equation}
\begin{aligned}
w &= \frac{1}{2 \pi \i}\int_0^{2 \pi} \d \theta_2 z(\theta_2) \dev{z^*(\theta_2)}{\theta_2} \\
& = \int_0^{2 \pi} \frac{\d \theta_2}{2 \pi \i} \left[ u^*(\theta_2+2 \pi \beta)\dev{u(\theta_2+2 \pi \beta)}{\theta_2} + u(\theta_2) \dev{u^*(\theta_2)}{\theta_2} \right] \\
& = \int_0^{2 \pi} \frac{\d \theta_2}{2 \pi \i} \left[ u^*(\theta_2)\dev{u(\theta_2)}{\theta_2} - u^*(\theta_2) \dev{u(\theta_2)}{\theta_2}  \right] \\
& = 0.
\end{aligned}
\label{eq:winding}
\end{equation}
A consequence of $w=0$ is that $\log z(\theta_2)$ is a smooth single valued imaginary function, from which we can (i) determine $u(\theta_2)$ and (ii) show $u(\theta_2)$ to be smooth. We find
\begin{align}
\label{eq:defu} 
u(\theta_2) &= \exp \left[\sum_{n\neq 0}\frac{f_n\e^{\i n \theta_2}}{1-\e^{2\pi \i \beta n}} \right],\\
f_n & = \frac{1}{2\pi}\int_0^{2\pi}\d \theta'\e^{-\i n \theta'}\log z(\theta')
\end{align}
which is verified to satisfy~\eqref{eq:zform} by substitution.
The smoothness of $u(\theta_2)$ follows as: (i) $\log z(\theta_2)$ is smooth, thus $f_n$ is exponentially small in $n$; (ii) the small values of the denominator of the summand are power law small in $n$~\cite{hindry2013diophantine},
\begin{equation}
|1-\e^{2\pi \i \beta n}| \gtrsim |F(\beta n)| > n^{-(\mu-1+\epsilon)}.
\end{equation}
Here $a_n\gtrsim b_n$ indicates $a_n > c b_n$ for some finite constant $c$ and sufficiently large $n$, $F(x) = x - \mathrm{nint}(x)$ is the fractional part of $x$, $\mathrm{nint}(x)$ is the nearest integer to $x$, $\epsilon$ is any arbitrarily small positive constant, and $\mu$ is the \emph{irrationality measure} of $\beta$. For the golden ratio, and almost all irrational numbers $\mu = 2$. The exponential smallness of $f_n$ beats the power law smallness of the denominator, and the terms of the sum in~\eqref{eq:defu} are exponentially small in $n$, and hence $u(\theta_2)$ is smooth.

As a result the state
\begin{equation}
\ket{\phi^j_\mathrm{M}(0,\theta_2)} = u^*(\theta_2)\ket{\phi^j_\mathrm{S}(0,\theta_2)} 
\end{equation}
defines a smooth gauge which has a constant phase difference on the boundary $\theta_1=0$:
\begin{equation}
\begin{aligned}
z(\theta_2) &=\bra{\phi^j_\mathrm{M}(0,\theta_2+2 \pi \beta)} U(T_1,0;0,\theta_2)\ket{\phi^j_\mathrm{M}(0,\theta_2)} \\
&= \e^{-\i \epsilon_j T_1},
\end{aligned}
\end{equation}
thus $\ket{\phi^j_\mathrm{M}(0,\theta_2)}$ satisfies the monodromy relation~\eqref{eq:monodromy}. 

As a final note we discuss the case of the topological class. In this analysis we used that $C_j=0$ implies $w=0$~\eqref{eq:winding}. In general one finds that the winding number $w = C_j$. When $w \neq 0$ no construction exists for a smooth $u(\theta_2)$, and no smooth gauge exists for which the monodromy relation is satisfied. However, instead, there is a smooth gauge that satisfies the generalised relation
\begin{equation}
U(T_1,0;0,\theta_2)\ket{\phi^j_\mathrm{S}(0,\theta_2)} = \mathrm{e}^{-\i (\lambda T_1 + C_j \theta_2)} \ket{\phi^j_\mathrm{S}(0,\theta_2+2\pi\beta)}.
\label{eq:polydromy}
\end{equation}

\section{Non-convergence of Floquet unitaries}
\label{app:FloqU}

\begin{figure}
\begin{center}
\includegraphics[width=0.4\textwidth]{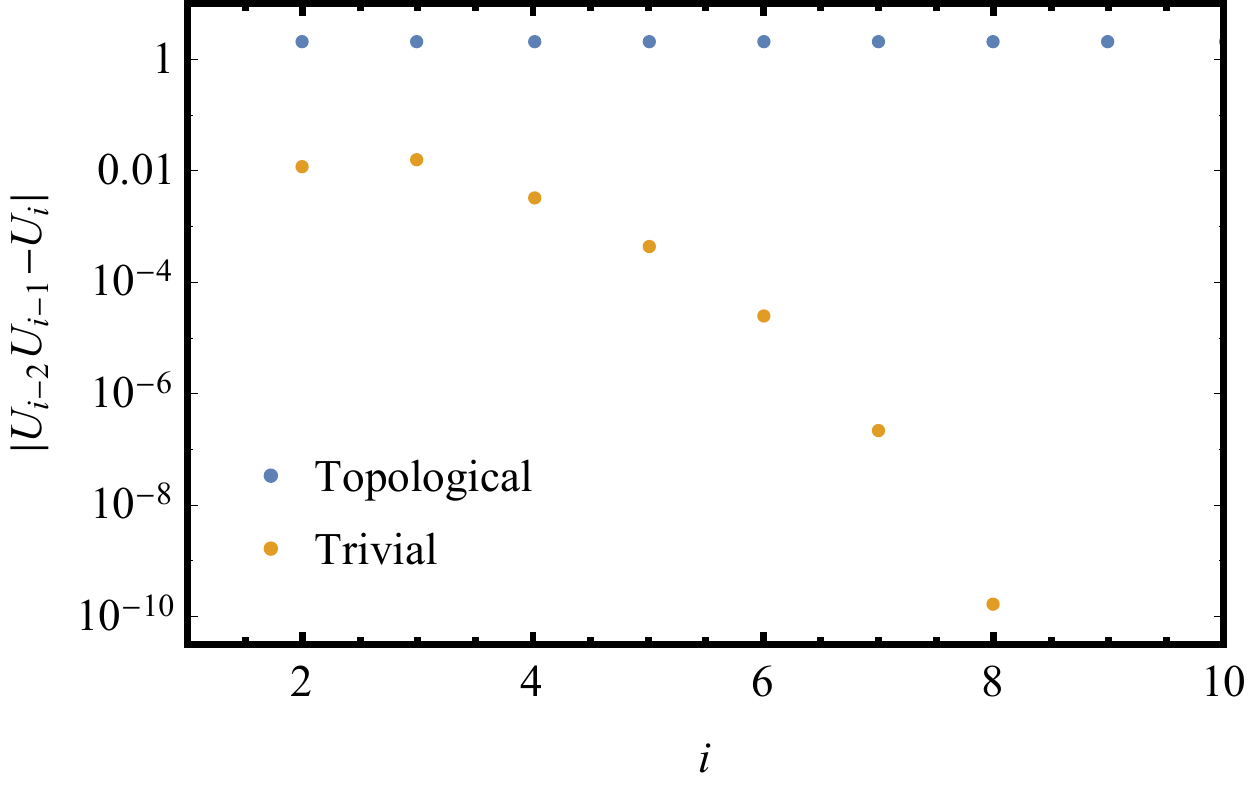}
\includegraphics[width=0.4\textwidth]{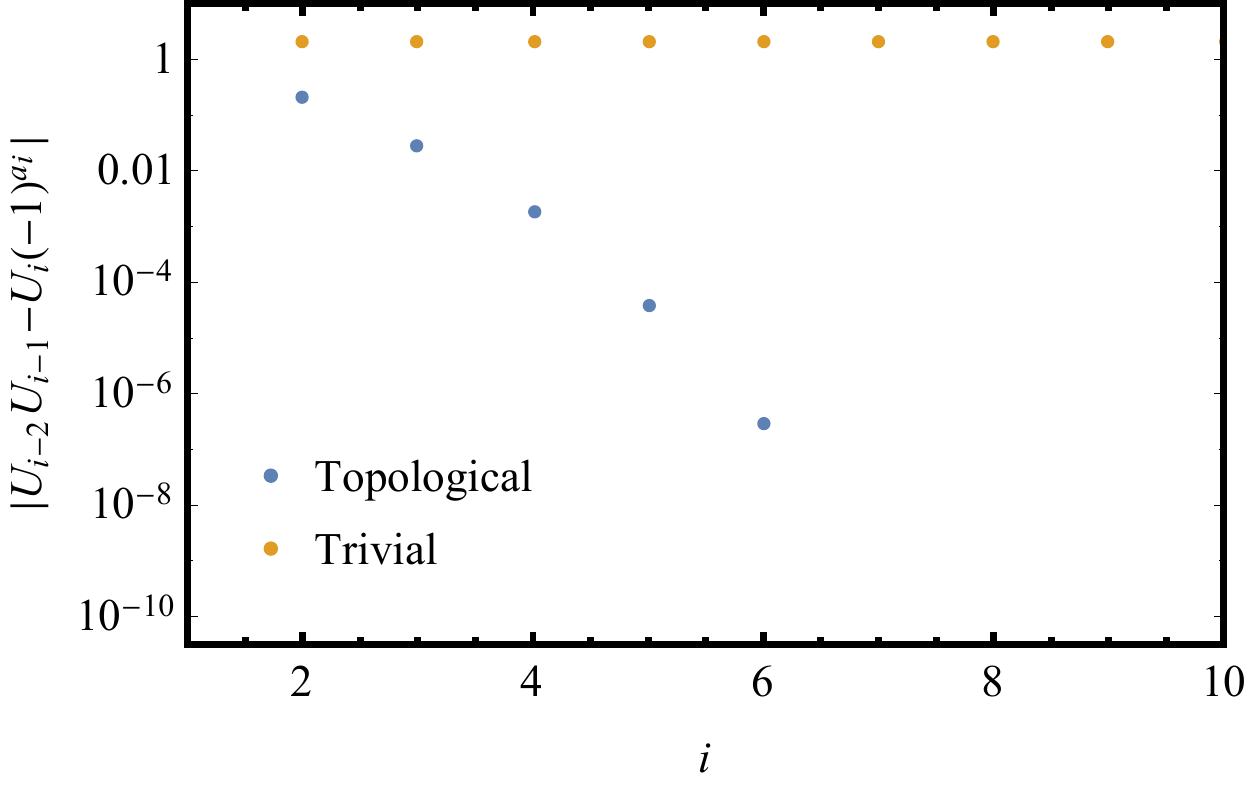}
\caption{
\emph{Convergence relation for Floquet unitaries:} The relation~\eqref{eq:UFtopocorr} is verified for the Floquet unitaries of $H_{\mathrm{CD}}$ in the topological and trivial cases. In the trivial phase, $C=0$ and the unitaries obey the relation $U_{i} \to U_{i-2}U_{i-1}^{a_{i}}$. In the topological case $C = \pm 1$ and the unitaries obey the relation $U_{i} \to (-1)^{a_{i}}U_{i-2}U_{i-1}^{a_{i}}$. In both cases the converging series is truncated when its value drops below numerical precision. Model $H_\mathrm{CD}$~\eqref{eq:topomodel} with parameters 
$T_1=4$, $T_2 = T_1 q_i/p_i$, $p_i/q_i=F_{i+2}/F_{i+1}$, $m=1(3)$ for topological(trivial), $\Bt_0=(0,0)$. }
\label{Fig:UF}
\end{center}
\end{figure}

In the main text we noted that in the trivial class of dynamics there is convergence of the Floquet unitaries of the dynamics in the commensurate approximnation to the time evolution operator of the incommensurate dynamics, whereas in the topological phase there is not~\eqref{eq:UiUt}. A numerically accessible consequence of this non-convergence is the convergence relation~\eqref{eq:UiUiUi} which relates successive Floquet unitaries calculated within the comensurate dynamics. In this appendix we derive the relation~\eqref{eq:UiUiUi}.

We first recall some properties of Diophantine approximation. For any irrational number $\beta$ the best rational approximations $p_i/q_i$ are given by truncating the continued fraction~\eqref{eq:cfrac}, or, equivalently, by the recursion relation 
\begin{align}
p_i &= a_i p_{i-1} + p_{i-2} \nonumber \\
q_i &= a_i q_{i-1} + q_{i-2}.
\label{eq:diophpq}
\end{align}
with initial values $p_{-1} = q_{-2} = 1$, $p_{-2} = q_{-1} = 0$. 
It is tempting to suggest that the Floquet unitaries may then satisfy the corresponding relation to~\eqref{eq:diophpq} given by
\begin{equation}
 U_{i} \ket{\phi^j(\Bt_0)} \xrightarrow{\text{?}} \left(  U_{i-1} \right)^{a_i} U_{i-2} \ket{\phi^j(\Bt_0)}.
\end{equation}
If the relation~\eqref{eq:ConvU} holds for all quasi-energy states it would imply that Floquet unitaries obey the following convergence
\begin{equation}
U_{i} \rightarrow \left(  U_{i-1} \right)^{a_i} U_{i-2}.
\label{eq:ConvU}
\end{equation}
We find that in the trivial case such a relation is satisfied,~\eqref{eq:ConvU} is verified numerically in the left Panel of Fig.~\ref{Fig:UF}. However non-trivial topology presents an obstruction to this convergence. 

We find however that~\eqref{eq:ConvU} is generically not satisfied in the topological case. To see this we note that the LHS of~\eqref{eq:ConvU} corresponds to the angular evolution
\begin{equation}
\binom{\theta_{t1}}{\theta_{t2}} = \binom{\Omega_1}{\Omega_1 p_i/q_i} t + \binom{\theta_{01}}{\theta_{02}} 
\label{eq:path1}
\end{equation}
whereas the RHS corresponds to the trajectory $\Bt_t'=\Bt_t + \dBt_t$
\begin{equation}
\binom{\delta\theta_{t1}}{\delta\theta_{t2}} = 
\begin{cases}
\binom{\Omega_1}{\Omega_1 (p_{i-2}/q_{i-2}-p_i/q_i)} t \quad & \text{ for } t < T_{i-2} \\
\binom{\Omega_1}{\Omega_1 (p_{i-1}/q_{i-1}-p_i/q_i)} t + \binom{0}{\Omega_1 (p_{i-2}/q_{i-2}-p_{i-1}/q_{i-1})} T_{i-2} \quad & \text{ for } T_{i-2}< t < T_{i} 
\end{cases}
\label{eq:path2}
\end{equation}
where $T_{i} = 2 \pi q_{i} / \Omega_1$. It is easily verified geometrically that the two trajectories define a triangle in the extended Floquet zone of area
\begin{equation}
\begin{aligned}
A &= \frac{1}{2} \left| \binom{\Omega_1 T_i}{\Omega_1 T_i  p_i/q_i} \times \binom{\Omega_1 T_{i-2}}{\Omega_1 T_{i-2} p_{i-2}/q_{i-2}} \right|
\\
&= \frac{1}{2} \left| \binom{2 \pi q_i}{2 \pi p_i} \times \binom{2 \pi q_{i-2}}{2 \pi p_{i-2}} \right|
\\
&= 2 \pi^2 a_i \left| \binom{q_{i-1}}{p_{i-1}} \times \binom{q_{i-2}}{p_{i-2}} \right|
\\
& =
2 \pi^2 a_i |p_{i-1}q_{i-2}-p_{i-2}q_{i-1}| 
\\ 
&= 2 \pi^2 a_i
\end{aligned}
\end{equation}
where $|p_{i-1}q_{i-2}-p_{i-2}q_{i-1}|=1$ is a standard result from the theory of Diophantine approximation.

Using the of Sec.~\ref{sec:divergence} that two quasi-energy state trajectories corresponding to slightly different trajectories through the Floquet zone separated by a small perturbation, accrue a phase difference $\e^{i \eta}$ where $\eta = A C_j / 2 \pi$ where $C_j$ is the Chern number of the quasi-energy state, and $A$ the area enclosed by the trajectories (in the case of Sec.~\ref{sec:divergence} $A= |\Bw||\dBt| t \sin \alpha$). In this case we thus find $\e^{i \eta} = \e^{\i \pi a_i C_j} = (-1)^{a_i C_j}$. In general, in addition to the correction to the phase, there is also a correction to the final state, however this correction goes to zero in the incommensurate limit $i \to \infty$ so we focus on the correction to phase. This implies the relation
\begin{equation}
 U_{i} \ket{\phi^j(\Bt_0)} \rightarrow (-1)^{a_i C_j} \left(  U_{i-1} \right)^{a_i} U_{i-2} \ket{\phi^j(\Bt_0)},
\label{eq:ConvU2}
\end{equation}
where the phase term constitutes a topological correction to the relation~\eqref{eq:ConvU}. Eq.~\eqref{eq:ConvU2} corresponds to~\eqref{eq:UiUiUi} in the main text. In the special case where $r=a_i C_j \mod 2$ is the same for all the partial quotients $a_i$, and the Chern numbers $C_j$ of all the quasi-energy-states, then the extra phase becomes a topological correction to~\eqref{eq:ConvU}
\begin{equation}
 U_{i} \to (-1)^{r}  U_{i-1}^{a_{i}} U_{i-2}.
\label{eq:UFtopocorr}
\end{equation}
this is verified numerically in Fig.~\ref{Fig:UF}.

\section{Divergence of trajectories}
\label{app:divergence}

In this appendix we derive the relation~\eqref{eq:errorgrowth}. This gives the coefficient of linear divergence of trajectories following a perturbation to the phases of the drives. This coefficient is zero in the trivial case, in which the trajectories do not diverge.

We begin by considering the states~\eqref{eq:psi1psi2}. These two trajectories define a parallelogram patch on the Floquet zone with vertices at $\Bt_0,\Bt_0+\dBt,\Bt_0+\Bw t,\Bt_0+\Bw t+\dBt$. We use~\eqref{eq:gauge_fix_phi} to fix the gauge choices between states related by shifts in the $\Bw$ direction, and choose a gauge such that $\ket{\phi^j(\Bt)}$ and $\epsilon_j(\Bt)$ are continuous in the perpendicular direction. As the quasi-energy state $\ket{\phi^j(\Bt)}$ define a smooth basis over the patch of interest the we can then resolve $\ket{\psi_1}$ in this basis. This yields
\begin{equation}
\ket{\psi_1(t)}=\sum_j \mathrm{e}^{-\i \epsilon_j(\Bt_0) t} \ket{\phi^j(\Bt_t)} \braket{\phi^j(\Bt_0)}{\psi_0}.
\end{equation}
Doing the same for $\ket{\psi_2(t)}$ and expanding in $\dBt$ yields
\begin{equation}
\ket{\psi_2(t)}=\ket{\psi_1(t)}+
\sum_j \mathrm{e}^{-\i \epsilon_j(\Bt_0) t} \left[ 
\ket{\partial_{\delta\Bt_0} \phi^j(\Bt_t)} \bra{\phi^j(\Bt_0)}
+
\ket{\phi^j(\Bt_t)} \bra{\partial_{\delta\Bt_0} \phi^j(\Bt_0)}
- \i t
\partial_{\delta\Bt_0} \epsilon_j(\Bt_0)\ket{\phi^j(\Bt_t)} \bra{\phi^j(\Bt_0)}
\right]\ket{\psi_0} + O(\delta\Bt_0)^2
\label{eq:psi2expand}
\end{equation}
where $\partial_{\dBt_0} = \dBt \cdot \nabla_{\Bt_0}$. 

In order to evaluate the distance between the states we use the infinitesimal form of the Bures angle
\begin{equation}
D_\mathrm{B}^2(\psi,\psi+\delta\psi) = \braket{\delta\psi}{\delta\psi} - \braket{\delta\psi}{\psi}\braket{\psi}{\delta\psi}
\label{eq:infBures}
\end{equation}
which can be verified by expansion of~\eqref{eq:BuresAngle}. We then see that if we evaluate the limit
\begin{equation}
\lim_{t\to\infty}\lim_{\delta\theta \to 0} \frac{1}{t^2|\dBt|^2} D_\mathrm{B}^2(\psi_1,\psi_2) 
\end{equation}
we will pick out the terms proportional to $(\partial_{\delta\Bt_0} \epsilon_j(\Bt_0))^2$ only, as this term is $O(t^2|\dBt|^2)$, and all other terms are higher order in $\dBt$ or lower order in $t$. Hence from~\eqref{eq:psi2expand}
and~\eqref{eq:infBures}
\begin{equation}
\lim_{t\to\infty}\lim_{\delta\theta \to 0} \frac{1}{t^2|\dBt|^2} D_\mathrm{B}^2(\psi_1,\psi_2)  = \sum_j \left[\partial_{\vec{u}} \epsilon_j(\Bt_0)\right]^2|\braket{\phi^j(\Bt_0)}{\psi_0}|^2 - \left[\sum_j \partial_{\vec{u}} \epsilon_j(\Bt_0)|\braket{\phi^j(\Bt_0)}{\psi_0}|^2\right]^2,
\end{equation}
where $\vec{u} = \dBt/|\dBt|$ is the unit vector parallel to $\dBt$ and $\partial_{\vec{u}} = \vec{u} \cdot \nabla_{\Bt_0}$.

From~\eqref{eq:grad_eps} it follows that 
\begin{equation}
\partial_{\delta\Bt_0} \epsilon_j(\Bt_0) = \frac{C_j}{2 \pi} (-\Omega_2,\Omega_1)\cdot (u_1,u_2) = \frac{C_j}{2 \pi}(\Omega_1,\Omega_2)\times (u_1,u_2) = \frac{C_j|\Omega|\sin{\alpha}}{2\pi}
\end{equation}
where $C_j$ is the Chern number of the quasi-energy state $\ket{\phi^j(\Bt)}$, $(\Omega_1,\Omega_2)\times (u_1,u_2) = u_2 \Omega_1-u_1\Omega_2 = |\Bw||\vec{u}|\sin \alpha$ is the antisymmetric product of a pair of two-element vectors, and $\alpha$ is the angle between $\Bw$ and $\dBt$. By defining $p_j = |\braket{\phi^j(\Bt_0)}{\psi_0}|^2$ this becomes
\begin{equation}
\lim_{t\to\infty}\lim_{\delta\theta \to 0} \frac{1}{t^2|\dBt|^2} D_\mathrm{B}^2(\psi_1,\psi_2)  = \left[\frac{|\Bw|\sin \alpha}{2\pi}\right]^2 \left[\sum_j C_j^2 p_j -  \left( \sum_j C_j p_j \right)^2 \right].
\end{equation}
Thus as $D_\mathrm{B}$ is defined to be positive we can invert the square on both sides without ambiguity.
\begin{equation}
\lim_{t\to\infty}\lim_{\delta\theta \to 0 } \frac{1}{t|\dBt|} D_\mathrm{B}(\psi_1,\psi_2) = \frac{|\Bw||\sin \alpha|}{2\pi} \sigma(C)
\end{equation}
yielding the form~\eqref{eq:errorgrowth} in the main text where $\sigma(C)$ is given by~\eqref{eq:sigmaC}.

\section{Scaling analysis of the topological and trivial regimes}
\label{app:Scaling}

\begin{figure}
\begin{center}
\includegraphics[width=0.8\columnwidth]{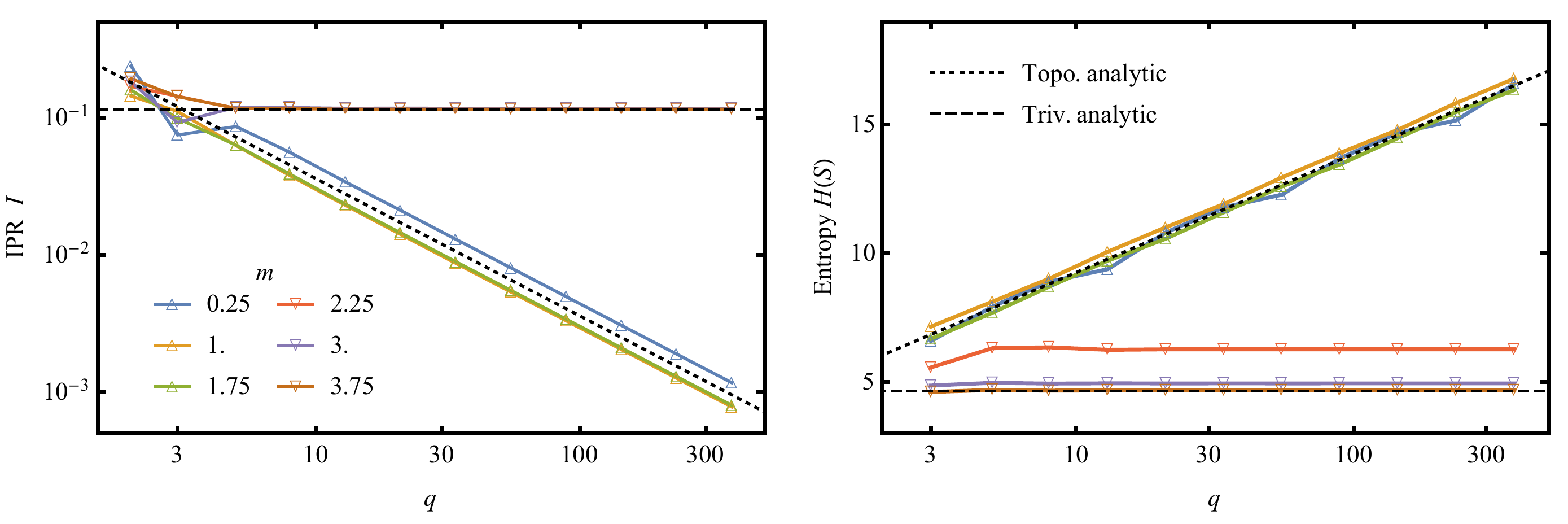}
\caption{
\emph{Localisation-delocalisation transition in frequency space of quasi-energy states and observables:} In the upper panel the inverse participation ratio $I$ (see~\eqref{eq:IPRdef}) of the quasi-energy-state is plotted versus $q$ for rational approximations~\eqref{eq:rational_drive}. Data from Hamiltonian~\eqref{eq:topomodel} which has a topological-trivial transition at $m=2$.
(Lower panel) the emergence of a region of dense spectrum in Fig.~\ref{Fig:OpSpec_fig} of topological origin is quantitatively verified. This is seen by the change in scaling with $q$ of $H(S)$ between the topological $(m<2)$ and trivial $(m>2)$ cases where $H(S) \sim \log q$ and $H(S) \sim \mathrm{cons.}$ respectively. Model $H_\mathrm{CD}$~\eqref{eq:topomodel} with parameters 
$T_1=4$, $T_2 = T_1 q_i/p_i$, $p_i/q_i=F_{i+2}/F_{i+1}$ for $q = F_3$ to $F_{14}$, $\Bt_0=(0,0)$. 
}
\label{Fig:IPR_fig1}
\end{center}
\end{figure}

In this appendix we numerically quantitatively verify two qualitative observations made in the main text: that, in the topological regime, the model~\eqref{eq:topomodel} gives rise to (1) quasi-energy states which are delocalised on the frequency lattice, (2) Fourier spectra of observables which are dense. These two properties are expected per Sec.~\ref{sec:deloc}. Here these properties are numerically verified using the scaling of the inverse participation ratio and the power spectral entropy (both statistics are defined below).

\subsection{Delocalisation on the frequency lattice}

Delocalisation can be verified quantitatively via the scaling of the inverse participation ratio (IPR)
\begin{equation}
I = \sum_\Bn \braket{\tilde\phi^j_\Bn(\Bt_0)}{\tilde\phi^j_\Bn(\Bt_0)}^2,
\label{eq:IPRdef}
\end{equation}
where $\ket{\tilde\phi^j_\Bn(\Bt_0)}$ are the Fourier components of the quasi-energy state $\ket{\phi^j(\Bt_0)}$ define in~\eqref{eq:QEdyn}. In the commensurate approximation, for delocalised states, $I \sim q^{-1}$, as each state is delocalised along a quasi-one-dimensional strip perpendicular to the electric field $\Bw$. Whereas in the localised case each state is restricted to a patch whose size is independent of the size of the lattice, and so $I$ does not scale with $q$.

This is shown for the quasi-energy-states of $H_\mathrm{CD}$ in Fig.~\ref{Fig:IPR_fig1} left panel. For the topological case $m<2$ (data points $\scriptstyle\triangle$) we see the delocalised behaviour $I \sim q^{-1}$ (guide line shown black dotted), whereas for $m>2$ we see $I \sim q^0$ (data points $\triangledown$) consistent with localised states. Data is shown for three values of $m$ in each of the topological and trivial case.

\subsection{Dense spectral entropy}

To establish the formation of a continuous part we investigate the behaviour \emph{power spectral entropy}
\begin{equation}
H(S) = -\sum_\omega s(\omega) \log s(\omega).
\end{equation}
where $s(\omega) = S(\omega)/\left[\sum_{\omega'} S(\omega')\right]$ is the \emph{normalised spectral density}, and $S(\omega)$ is defined in~\eqref{eq:PSobs}.

Data for $H(S)$ calculated within the commensurate approximation is shown in the right panel of Fig.~\ref{Fig:IPR_fig1}. For a pure point spectrum we expect there are finitely many contributions to the sum in $H(S)$ as $q \to \infty$ and we expect $H(S) \sim \mathrm{cons.}$, this is seen for data from the trivial case (data points $\triangledown$). Whereas, in the topological case (data points $\scriptstyle\triangle$), there are infinitely many contributions to the sum coming from the formation of the dense part of the spectrum, and $H(S) \sim \log q$.

\section{Averaged power spectrum}
\label{app:Sw}

Here we give the form of the averaged power spectrum used to make Fig~\ref{Fig:OpSpec_fig}, which is obtained by analytically performing the averages of~\eqref{eq:PSobs}. From~\eqref{eq:FTobs} we have that the Fourier spectrum $A(\omega)$ of the expectation value of a generic operator is given by
\begin{equation}
A(\omega) = 
\begin{cases}
\sum_{\epsilon}\sum_{\Bm} \alpha_j^*\alpha_j \bra{\phi_{\Bm}^j} A \ket{\phi_{\Bn+\Bm}^{j}}\, & \mathrm{for} \quad \, \omega = \Bw \cdot \Bn\\
\sum_{\Bm} \alpha_j^*\alpha_{j'} \bra{\phi_{\Bm}^j} A \ket{\phi_{\Bn+\Bm}^{j'}}\, & \mathrm{for} \quad \, \omega = \Bw \cdot \Bn + \epsilon_{j'}-\epsilon_j\\
\sum_{\Bm} \alpha_{j'}^*\alpha_j \bra{\phi_{\Bm}^{j'}} A \ket{\phi_{\Bn+\Bm}^{j}}\, & \mathrm{for} \quad \, \omega = \Bw \cdot \Bn + \epsilon_j-\epsilon_{j'}\\
\end{cases}
\end{equation}
Here, $j\neq j'$, and we define the $A^{j,j'}_{\Bn} = \sum_\Bm \bra{\phi_{\Bm}^j} A \ket{\phi_{\Bn+\Bm}^{j'}}$ which can be obtained numerically by Fourier transform of $A$ sandwiched between quasi-energy states (which are in turn obtained from numerical integration of the equations of motion) as follows
\begin{equation}
\bra{\phi^j(\Bt_t)}A\ket{\phi^{j'}(\Bt_t)} = \sum_{\Bn \Bm} \bra{\phi_{\Bm}^j} A \ket{\phi_{\Bn+\Bm}^{j'}} \mathrm{e}^{- \i \Bw \cdot \Bn t} = \sum_\Bn A^{j,j'}_{\Bn} \mathrm{e}^{- \i \Bw \cdot \Bn t} 
\end{equation}
Using this, and Eqn.~\eqref{eq:FTobs}, we find this quantity is related to the Fourier spectrum by
\begin{equation}
A(\omega) = 
\begin{cases}
|\alpha_j|^2 A^{j,j}_{\Bn} + |\alpha_{j'}|^2 A^{j',j'}_{\Bn}\, & \mathrm{for} \quad \, \omega = \Bw \cdot \Bn\\
\alpha_j^*\alpha_{j'} A^{j,j'}_{\Bn}\, & \mathrm{for} \quad \, \omega = \Bw \cdot \Bn + \epsilon_{j'}-\epsilon_j \\
 \alpha_{j'}^*\alpha_j A^{j',j}_{\Bn} \, & \mathrm{for} \quad \, \omega = \Bw \cdot \Bn + \epsilon_j-\epsilon_{j'}\\
\end{cases}
\end{equation}
where we have assumed $j \neq j$ and the non-degeneracy of the quasi-energies $\epsilon_j \neq \epsilon_{j'}$. Integrating over the Bloch sphere we find there are four non-zero contributions from the sum over the quasi-energies: two from $\left[ |\alpha_j|^4 \right]_{\ket{\psi_0}} = \left[ |\alpha_{j'} |^4 \right]_{\ket{\psi_0}} = 2/3$; and two from $\left[ |\alpha_j|^2 |\alpha_{j'} |^2 \right]_{\ket{\psi_0}} = 1/3$. This yields
\begin{equation}
\left[S(\omega)\right]_{\ket{\psi_0}} = 
 \begin{cases}
\frac{2}{3} \left( 
|A^{j,j}_{\Bn}|^2
+
|A^{j',j'}_{\Bn}|^2
+
|A^{j,j}_{\Bn}| |A^{j',j'}_{\Bn}|
\right) 
\, & \mathrm{for} \quad \, \omega = \Bw \cdot \Bn\\
\frac{1}{3} |A^{j,j'}_{\Bn}|^2 & \mathrm{for} \quad \, \omega = \Bw \cdot \Bn + \epsilon_{j'}-\epsilon_j \\
\frac{1}{3} |A^{j',j}_{\Bn}|^2  & \mathrm{for} \quad \, \omega = \Bw \cdot \Bn + \epsilon_j -\epsilon_{j'}.
\end{cases}
\end{equation}
The integration over $A = \vec{a} \cdot \vec{\sigma}$ is realised using the relation $\left[ a_\alpha a_\beta \right]_A = \frac{1}{3}\delta_{\alpha,\beta}$ where the $\left[ \cdot \right]_A$ denotes the uniform integration $\int_{S^2} \cdot \mathrm{d}^2 \vec{a}$ over the unit sphere $|a|=1$. This yields
\begin{equation}
\left[S(\omega)\right]_{\ket{\psi_0},A} = 
 \begin{cases}
\frac{2}{9} \sum_{A \in \vec{\sigma} } \left( 
|A^{j,j}_{\Bn}|^2
+
|A^{j',j'}_{\Bn}|^2
+
|A^{j,j}_{\Bn}| |A^{j',j'}_{\Bn}|
\right) 
\, & \mathrm{for} \quad \, \omega = \Bw \cdot \Bn\\
\frac{1}{9} \sum_{A \in \vec{\sigma} }  |A^{j,j'}_{\Bn}|^2 & \mathrm{for} \quad \, \omega = \Bw \cdot \Bn + \epsilon_{j'}-\epsilon_j \\
\frac{1}{9} \sum_{A \in \vec{\sigma} }  |A^{j',j}_{\Bn}|^2  & \mathrm{for} \quad \, \omega = \Bw \cdot \Bn + \epsilon_j -\epsilon_{j'} \\
\end{cases}
\end{equation}
where $\vec{\sigma}$ is the usual vector of Pauli matrices. This is the equation used to obtain the Figures used in the main text and App.~\ref{app:Scaling}.

\end{widetext}

\end{document}